\documentclass{aa}

\usepackage{epsfig,graphicx,natbib}
\usepackage{longtable}
\usepackage{amssymb}
\usepackage{amsmath}
\usepackage{xcolor}
\usepackage{booktabs}

\usepackage{bibentry}
\usepackage{lineno}

\newcommand{\ignore}[1]{}
\newcommand{\nobibentry}[1]{{\let\nocite\ignore\bibentry{#1}}}

\newcommand{\lsim}{\raise0.3ex\hbox{$<$}\kern-0.75em{\lower0.65ex\hbox{$\sim$}}}
\newcommand{\gsim}{\raise0.3ex\hbox{$>$}\kern-0.75em{\lower0.65ex\hbox{$\sim$}}}
\newcommand{\propsim}{\raise0.3ex\hbox{$\propto$}\kern-0.75em{\lower0.65ex\hbox{$\sim$}}}

\def\SN93J{SN\,1993J}

\begin{document}

\title{
 The radial distribution of radio emission from \SN93J: Magnetic field amplification due to the Rayleigh-Taylor instability
}

\author{ I. Mart\'i-Vidal \inst{1,2} 
\and C-I. Bj\"ornsson\inst{3} 
\and M.~A. P\'erez-Torres\inst{4,5}
\and P. Lundqvist\inst{3,6} 
\and J.~M. Marcaide\inst{7}
}

\institute{
  Departament d’Astronomia i Astrof\'isica, Universitat de Val\`encia, C. Dr. Moliner 50, 46100 Burjassot ,Val\`encia, Spain
  \and  
  Observatori Astron\`omic, Universitat de Val\`encia, Parc Cient\'ific, C. Catedr\`atico Jos\'e Beltr\'an 2, 46980 Paterna, Val\`encia, Spain
  \and
  Department of Astronomy, AlbaNova, Stockholm University, 106 91 Stockholm, Sweden
   \and  
  Instituto de Astrof\'isica de Andaluc\'ia (CSIC), Glorieta de la Astronom\'ia s/n, 18008 Granada, Spain
  \and
     School of Sciences, European University Cyprus, Diogenes street, Engomi, 1516 Nicosia, Cyprus
  \and
  The Oskar Klein Centre, AlbaNova, Stockholm University, 106 91 Stockholm, Sweden 
  \and
  Real Academia de Ciencias Exactas, F\'isicas y Naturales de Espa\~na,
Calle Valverde 22, 28004 Madrid, Spain
}

\date {Received  / Accepted}

\titlerunning{Inhomogeneous radial distribution in the \SN93J radio emission}
\authorrunning{Mart\'i-Vidal et al.}

%\LEt{***General notes: (A.) I have edited to US English spelling and grammar conventions. (B.) A&A uses the past tense to describe specific methods used in a paper and the present tense to describe general methods as well as findings, including the findings of recent papers (within the past ten or so years). Kindly make any necessary changes (I have made some, but my edits are by no means exhaustive in this respect). See Sect. 6 of the language guide https://www.aanda.org/for-authors/language-editing/6-verb-tenses}
\abstract
%% CONTEXT
{Observations of radio emission from young core-collapse supernovae (CCSNe) allow one to study the history of the pre-supernova stellar wind, trace the density structure of the ejected material, and probe the magnetohydrodynamics that describe the interaction between the 
two, as the forward shock expands into the circumstellar medium. The radio shell of supernova SN\,1993J has been observed with very long baseline interferometry (VLBI) for $\sim$20 years, giving one of the most complete pictures of the evolution of a CCSN shock. However, different results about the expansion curve and properties of the radio-emitting structure have been reported by different authors, likely due to systematics in the data calibration and/or model assumptions made by each team.}
%% AIMS
{We aim to perform an analysis of the complete set of VLBI observations of SN\,1993J that accounts for different instrumental and source-intrinsic effects, in order to retrieve robust conclusions about the shock expansion and physics in SN\,1993J.}
%% METHODS
{We have explored the posterior probability distribution of a complete data model, using a technique based on Markov chains. Our model accounts for antenna calibration effects, as well as different kinds of radio-emission structures for the supernova.}
%% RESULTS
{The posterior parameter distributions strongly favor a spherical shell-like radio structure with a nonuniform radial intensity profile, with a broad brightness distribution that peaks close to or just above the region where the contact discontinuity is expected to be located.
Regarding the shell expansion, the full dataset can be well described using one single deceleration parameter, $\beta \sim 0.80$, being the shell outer radius $R \propto t^\beta$. There is clear evidence of a relative widening of the shell width beyond day 2600$-$3300 after the explosion, which is due to an increased deceleration of the inner shell boundary. This is similar to findings previously reported by other authors.}
%% CONCLUSIONS
{The radial intensity profile and the late evolution of the shell suggest a scenario in which the magnetic field is amplified mainly by the Rayleigh-Taylor instability, which emanates from the contact discontinuity. Furthermore, the increased deceleration of the inner boundary indicates that the reverse shock enters a region of the ejecta at around 3000 days, where the density distribution is substantially flatter. Such a weakening of the reverse shock can also explain the achromatic break in the radio light curves, which occurs at the same time. The deduced radial intensity distribution for SN\,1993J is quite similar to that observed in the spatially well-resolved supernova remnant Cassiopeia A.
}

\maketitle

\section{Introduction}

SN\,1993J is arguably one of the most well-observed radio supernovae, together with SN\,1987A, not only because of detailed spectra over a long period of time but also, due to its proximity, for its spatially resolved source structure. Therefore, it is a prime object to which the predictions of various theoretical models can be compared. The evolution of its radio emission during the first few hundred days was nonstandard in that the peak spectral flux increased with time and very long baseline interferometry (VLBI) observations showed the source to expand at a roughly constant velocity. After this initial phase, the evolution transitioned to a more standard behavior with a roughly constant spectral peak flux and a decreasing rate of expansion \citep{wei07,Marcaide2009,bie11}. It has been argued that the initial rise of the peak spectral flux was caused by synchrotron cooling \citep{f/b98,per01}. This requires a strong magnetic field, and hence a high thermal energy density in between the forward and reverse shocks. Since the shocked region is expected to follow a self-similar evolution \citep{che82}, the roughly constant velocity of the forward shock indicated a very steep density gradient of the supernova ejecta.

However, as is discussed in \cite{bjo15}, the synchrotron cooling scenario has several implications that are at odds with observations; in particular, the details of the transition to the later, more standard evolution. Furthermore, a comparison to explosion models developed to account for the optical light curves of SN\,1993J shows that the kinetic energy of the ejecta required in the synchrotron cooling scenario is more than an order of magnitude larger than even the most favorable model can provide. Another observation arguing against an initial steep density gradient of the ejecta is the evolution of the X-ray light curves. As was pointed out by \cite{fra96}, a gradient strong enough to explain the roughly constant initial velocity would also imply cooling behind the reverse shock. Thus, a thin shell of cold gas would form and absorb most of the emitted X-ray emission. Since the X-ray emission is expected to be dominated by the reverse shock, an initially declining X-ray light curve would reverse when the cold shell becomes optically thin. However, no such increase was observed; not even a leveling of the X-ray light curves \citep{cha09}. 

Instead, it was suggested in \cite{bjo15} that the initial phase of the radio emission in SN\,1993J resulted from the Rayleigh-Taylor instability at the contact discontinuity, which is driven by the reverse shock. This is in contrast to the standard model, where both the acceleration of the relativistic electrons and the amplification of the magnetic field are due to the forward shock. As was shown in \cite{che92}, the turbulence resulting from the Rayleigh-Taylor instability grows outward from the contact discontinuity until it saturates close to the forward shock. The turbulence is expected to amplify the magnetic field \citep{j/n96a,j/n96b}. The initial phase in SN\,1993J would then be the combination of an expanding emission region together with a decelerating forward shock, resulting in a roughly constant velocity of its outer boundary.

One may note that such a model is similar to the resolved structure observed in Tycho/SN\,1572 \citep{dic91}. Here, most of the radio emission comes from an inner region bounded by the reverse shock and only a small fraction from a distinct but thin outer shell associated with the forward shock. A very similar emission structure is also found in Cassiopeia A \citep{got01}. Since the synchrotron emission region is defined by its magnetic field, the site for the acceleration of the relativistic electrons could be either the forward shock, or the turbulent region amplifying the magnetic field.

At around 3000\,days, a smooth but distinct achromatic break occurred in the radio light curves \citep{wei07}, which coincided with a similar decline in the X-ray light curves \citep{cha09,dwa14}. This event also affected the optical spectrum of SN\,1993J. Shortly before 400\,days, it had transitioned into a phase characterized by box-like emission line profiles and, in particular, H$\alpha$ had become dominant. As was discussed by \citep{mat00a}, this suggested the emission lines originated in an optically thin, spherically symmetric shell and that circumstellar interaction provided the energy input. This phase lasted til around day 2454, when the dominance of H$\alpha$ started to subside \citep{mat00b}. A few years later the box-like profiles had gone completely and the dominant emission line was instead  [O\,III] \citep{mil12}. 

The X-ray emission as well as the box-like emission line profiles are thought to come from the region associated with the reverse shock \citep{c/f94}. The cause of their simultaneous decline could then be a weakening of the reverse shock. If so, this suggests that the radio emission is also determined mainly by the strength of the reverse shock, analogous to the situation in Tycho and Cassiopeia\,A. 

Such a scenario is supported by the VLBI observations of \cite{bie11}. After a few hundred days, the outer boundary of the radio emission region started to increase with time as $\sim t^{0.8}$. When the inner and outer boundaries could be followed independently, they showed the same expansion, as was expected for a self-similar structure of the emission region \citep{che82}. However, at around 3000\,days this behavior changed. The inner boundary started to lag behind — its deceleration increased — while the outer boundary seemed to continue unaffected. This slowdown may be due to a decrease in the momentum input to the reverse shock from the ejecta, which would also weaken the shock. Hence, this observation adds strength to the argument that the radio emission, just like the X-ray emission and the optical line emission, is directly related to the physical conditions behind the reverse shock. The decrease in the momentum input would indicate that the reverse shock had entered a region of the ejecta with a smaller density gradient. This is likely associated with the transition from the hydrogen to the helium dominated shell of the ejecta. As is pointed out in \cite{bjo15}, the deduced ejecta velocity at which this transition occurs fits well with the ejecta structure in the preferred model for the optical light curves; that is, model 13B of \cite{woo94}.

The VLBI observations are the most direct and model-independent evidence for a weakening reverse shock as the common cause of a range of observed changes in SN\,1993J taking place at around 3000\,days. It is therefore essential not only to confirm the observational results of \cite{bie11} but also to find indications for other effects expected to be associated with such a weakening of the reverse shock. A decreasing momentum input should also cause a slowing down of the forward shock. However, this is likely to occur over a longer timescale than that for the reverse shock. Furthermore, the increased deceleration is not expected to be that strong, since the observed expansion ($t^{0.8}$) is not that different from a pure Sedov expansion ($t^{2/3}$), which corresponds to zero momentum input; that is, no reverse shock. The long timescale associated with changes at the forward shock can be seen in the hydrodynamic simulations done in \cite{kun19}. 

One important aspect of the VLBI observations not discussed in \cite{bie11} is the radial distribution of the radio emission in between the forward and reverse shocks. This is an important aspect for the scenario of a weakening reverse shock, since it directly addresses the amplification of the magnetic field and the acceleration of the relativistic electrons. Although the strength of the magnetic field is expected to be determined by the turbulence driven by the Rayleigh-Taylor instability, the mechanism for accelerating the electrons is less clear. As was mentioned already, it could be related to the turbulence but it is also possible that the electrons are energized through standard first-order Fermi acceleration at the forward shock. Of particular importance for an understanding of the physical processes operating in the inter-shock region would be to observe changes in the radial distribution of the synchrotron emission caused by a weaker reverse shock.

In this paper, we present a re-analysis of the VLBI observations of SN\,1993J, focusing on the relevant issue of the radial intensity profile of the shell. To this end, we apply a Markov chain Monte Carlo (MCMC) technique, which 
takes into account the biasing effects related to the coupling of the radial intensity with other source parameters, as well as with the limitations in the data calibration and modeling. To our knowledge, this is the first time that such an approach has been applied in VLBI studies of supernovae.  

In Sect. \ref{ObsCalSec}, we present the set of observations, the calibration procedure, and the modeling methodology.
In Sect. \ref{ResultSec}, we summarize the results and describe different tests performed to the data, in order to quantify the possible sources of parameter bias (in particular, the use of different weighting schemes for the visibilities, the possible presence of inhomogeneities in the shell azimuthal brightness distribution, and possible effects related to the radial intensity profile). 

In Sect. \ref{DiscussSec}, we compare our results to those from previous publications; in particular, we discuss the SN expansion rate and its shell width, as well as the wavelength effects in both of them.  
In Sect. \ref{DiscussSec2}, we discuss the implications of our results on the radio emission model of radio supernovae and reflect on the problems intrinsic to the use of VLBI observations in the modeling of the radio emission from supernovae.

\section{Data calibration and modeling}
\label{ObsCalSec}

\SN93J was observed with the Very Long Baseline Array (VLBA) and other global VLBI arrays, including the European VLBI Network (EVN) shortly after its discovery on 28 March 1993 \citep{Ripero1993}. The high latitude of its host galaxy, M\,81, made \SN93J a suitable VLBI target at any epoch of the year. The angular proximity of the active galactic nucleus (AGN) of M\,81 ensured good detections for phase-referencing calibration. Those facts have made it possible to accumulate more than 60 VLBI observing epochs at several frequencies (mainly from 1 to 10 GHz) over almost twenty years. With the exception of SN\,1987A, supernova 1993J is, by far, the most and best observed extra-galactic radio supernova.

Lists of available VLBI epochs of \SN93J are given in the several works that have been published so far \citep[e.g., ][]{Bietenholz2001, Marcaide2009, MartiVidal2011a}. For details about the dates, integration times, observation strategies, participating antennas, and frequency configurations for each epoch, we refer the reader to all these previous publications. 

In this work, we make a combined use of all the VLBI observations available for \SN93J that were phase-referenced to the AGN in M81, which excludes some of the very early observations (a list of the epochs used in this analysis is given in Appendix \ref{app:Table}). There is, actually, one extra epoch at the L band (project GB070, centered at 1.66\,GHz) observed on March 5, 2010 that has not yet been reported in any refereed publication \citep[although the image was published in a conference proceedings by ][]{bie11}. We have also included this epoch (the very last VLBI observations of \SN93J, to our knowledge) in our analysis. This paper thus represents the most complete VLBI work on \SN93J.

\subsection{The shell brightness distribution}

According to the model of \cite{Chevalier1982}, the brightness distribution of \SN93J at radio wavelengths can be described as a spherical shell, with an outer radius, $R_o$, related to the shocked circumstellar medium (CSM) and an inner radius, $R_i$, likely related to the shocked ejecta. The brightness distribution across the radio shell has been so far assumed to be constant in the modeling of the VLBI observations of \SN93J \citep[e.g.,][]{Bartel2002, Marcaide2009}, which implicitly assumes an homogeneous distribution of the relativistic electrons and magnetic-field intensity in the shocked CSM region.

The incomplete coverage of the Fourier (UV) space by the VLBI arrays and the limited signal-to-noise ratio (S/N) of the \SN93J visibilities have an important impact on the estimates of the shell size and width. On one hand, using fitting models with shell widths that differ from the true width of the supernova remnant shell may result in running biases on the estimated shell size \citep{Marcaide1997,Marcaide2009, MartiVidal2011a}, which affect the inferred supernova expansion rate. Such kinds of biases may also change, depending on whether the shell expansion is not self-similar (e.g., if the relative shell width, its radial intensity profile, and/or the internal opacity due to synchrotron self-absorption or free-free absorption, evolve).

The use of adaptive UV tapers — that is, the decrease in the statistical weights of the longest interferometric baselines — helps to decrease the effects of running biases that may be related to both the expanding structure and the finite sensitivity of the interferometer \citep[e.g.][]{Marcaide2009}. However, the choice of the UV-taper size and scaling is rather arbitrary. In addition to this, the use of classical hybrid imaging algorithms \citep[e.g.][]{HybridRef} may result in spurious contributions to the source structure \citep[e.g.,][]{MartiVidalSpur}, which would have unpredictable effects on the fit shell parameters.

All the possible issues summarized in the previous paragraphs contaminate the estimates of the shell size and width. Actually, the different 
shell parameters and expansion curves of \SN93J that have been reported so far by different authors (see Sect. \ref{DiscussSec} for a more detailed discussion) may be explained by different choices of the model-fitting structures, visibility weighting schemes, and calibration approaches.

In this paper, we study the expansion and evolving shell structure of \SN93J by quantifying the most important effects that contribute to the visibility calibration and fitting.  Namely, the amplitude calibration of the radio telescopes \citep[with typical relative deviations of the order of 10-20\%, especially for heterogeneous arrays, e.g.][]{Bondi1994}, the phase-referencing astrometry shifts, and the different shell-structure assumptions. The dimensionality of the resulting parameter space, and the tools used for its complete exploration, is discussed in the following subsections.

\subsection{Modeling based on Markov chains}
\label{MethodSec}

We have analyzed the shell structure (position, size and width) and effects from the antenna calibration, by carrying out a complete exploration of the parameter space with Markov chains \citep[e.g.][]{MarkovRef}, using a Metropolis-based Monte Carlo approach (MCMC). The use of MCMC allows us to explore parameter spaces of models with a large number of free parameters, with the advantage of probing regions with different local minima of the $\chi^2$ (i.e., regions that other methods, based on $\chi^2$ gradients, would fail to properly sample). The parameters used in our MCMC models are:

\begin{enumerate}
    \item Shell position offset, $(\Delta\alpha, \Delta\delta)$, with flat priors of 10\,mas with respect to the (phased-referenced) supernova center. These parameters would account for possible phase-calibration shifts related to the (changing) relative astrometry between \SN93J and its calibrator, the low-luminosity AGN in M\,81 (results from this relative astrometry, and its interpretation in terms of the jet physics of the AGN in M\,81, can be found in \citealt{BietenholzM81} and \citealt{MartiVidalM81}).   
    \item Total shell flux density, $S_{tot}$, with a flat positive prior (i.e., no negative flux densities are allowed).
    \item Shell outer radius, $R_{o}$, with a flat positive prior (i.e., no negative sizes are allowed).
    \item Shell inner radius, $R_{i}$, with a flat prior between 0 and $R_{o}$. We apply this prior by setting a flat prior to the relative size ratio, $R_{i}/R_{o} \in [0,1]$.
    \item Radial intensity slope of the shell brightness, $R_{sl}$ ( with a flat prior $\in [-1,1]$) or peak relative radial brightness position, $\rho_{peak}$ (with a flat prior between $R_i/R_o$ and 1). These parameters are only used for the analysis discussed in Sect. \ref{RadialProfileSec}.
    \item Global amplitude-gain corrections for each antenna, with flat positive priors (i.e., no negative amplitude gains are allowed).
\end{enumerate}

There are some important differences between our MCMC approach and the analyses of the \SN93J expansion reported in the previous publications. First, the MCMC chains provide the posterior distributions of all the model parameters, so that the derived uncertainties reflect the behavior of the $\chi^2$ in a more realistic, unbiased way. Methods based in the linearization of the $\chi^2$ gradient (like the Levenberg-Marquardt, e.g., \citealt{UVMultiFit}) usually provide uncertainties based on the post-fit Hessian matrix, which only accounts for the local behavior of the $\chi^2$ around the minimum (or one of the possible minima).

Second, we include the effect of the antenna gains in the SN modeling, so that it accounts for the cross-talk between the supernova structure and possible instrumental (calibration) effects. In this model, the phase gains are taken from the phase referencing with respect to M\,81*, with no further self-calibration. The relative changes in the antenna amplitude gains are obtained from the system temperatures and gain curves of each station. We note that in our MCMC model we allow for global scaling of the amplitude gains of each antenna, which (based on our experience) may depart from unity even if the system temperatures are properly applied to the data. The cause for these antenna global amplitude factors is not clear, but could be due to, for example, pointing errors, or phase noise in the local oscillator, and may typically affect the amplitude calibration by factors as large as 10$-$20\%.

Simultaneous modeling of source structure and instrumental gains in VLBI, using MCMC approaches, have already been reported in several publications \cite[e.g.,][]{VLBIMCMC,ThemisRef,DMCRef}, where their advantages over classical (hybrid imaging) algorithms are discussed. Here, it suffices to say that the inclusion of antenna gains in the MCMC chain yields more realistic uncertainties of the supernova shell structure, as well as of the shell parameter estimates, where possible gain-related biases \citep[e.g., the gains of antennas commonly appearing in longer baselines, like MK or SC; see][for a discussion]{MartiVidalResol} may affect the results. 

Furthermore, the use of gradient-based $\chi^2$ minimization methods for parameter fitting relies on the linearization of the $\chi^2$ gradient around the (possibly nonunique) minimum. This implies that the probability distributions of the fit parameters are assumed to be Gaussian. Any deviation from linearity in the $\chi^2$ gradient results in non-Gaussian parameter distributions, which may show, for example, skewness in some directions of the parameter space. All these effects are properly treated by MCMC methods.

\begin{figure}[ht!]
\includegraphics[width=9.5cm]{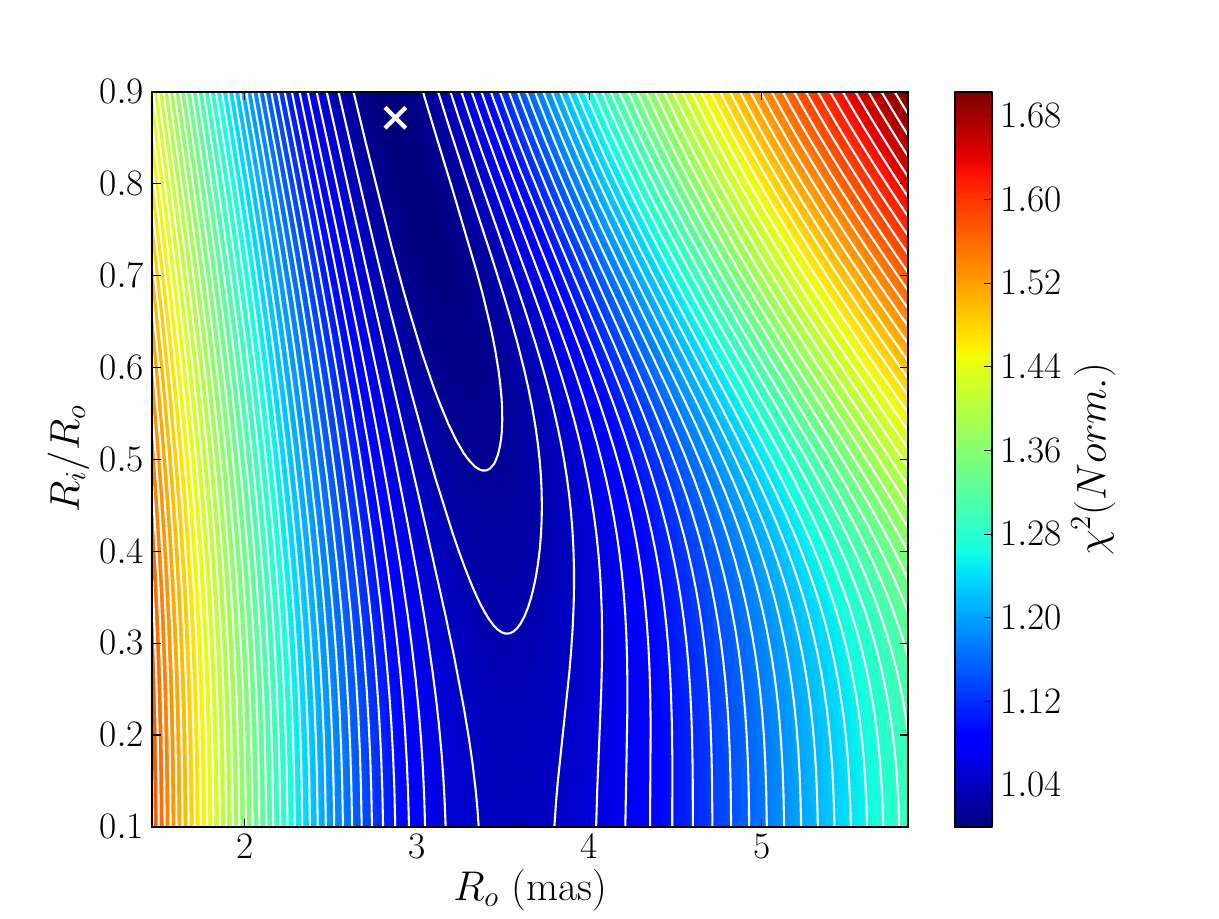}
\caption{$\chi^2$ of a shell fit to the data taken on day 1639 at C band (5\,GHz), as a function of outer radius (horizontal axis) and ratio of inner-to-outer radius (vertical axis). The $\chi^2$ values are normalized to the global minimum and the total shell flux density is optimized at each point. The coordinates of the shell center are fixed to the estimates from a fit with all free parameters.}
\label{Chi2DFig}
\end{figure}

As an example of non-Gaussian parameter behavior, we show in Fig. \ref{Chi2DFig} the $\chi^2$ distribution for a 2D slice of the parameter space of shell model, formed by the outer radius, $R_{o}$ (horizontal axis), and the size ratio (i.e., inner-to-outer radii, $R_i/R_o$, vertical axis), fit to the \SN93J data taken at 5\,GHz on day 1639 after explosion (21 September 1997). All the other model parameters are set to their optimum values (i.e., those that minimize the $\chi^2$). The shape of the $\chi^2$ in the direction of $R_i/R_o$ departs from a paraboloid (i.e., its gradient is far from linear), with a steep and narrow valley that actually resembles the Rosenbrock function \citep{RosenRef}. Apart from the limitations that a pure gradient-based method may encounter while minimizing the $\chi^2$ along that valley, the size ratio may have a wide and asymmetric probability density distribution, which is not properly modeled by a simple Gaussian.

\subsection{Procedure for the Markov chain Monte Carlo exploration}

The complete set of VLBI observations of \SN93J is very heterogeneous. The gain curves and receiver efficiencies of many of the participating antennas changed (improved) several times during the almost 20 years of duration of the whole SN\,1993J campaign. In addition to this, the number of participating antennas is highly variable across the campaign. There are epochs of global (VLBA + EVN), very sensitive observations, along with many other close-by epochs having a much more limited UV coverage. Last but not least, the frequency configurations, observation strategies (e.g., different phase-referencing duty cycles, amplitude calibrators, etc.) and even the correlation coordinates, were different between the two teams that observed SN\,1993J \citep[for details, see e.g.,][]{Marcaide2009,Bartel2002}.

A homogeneous MCMC analysis of such a heterogeneous dataset is not simple. The exploration of the posterior probability distributions of the parameters relies on a correct scaling of the visibility uncertainties, which may be very different across time and frequency. The uncertainties of the visibilities are scaled based on the amplitude calibration, combined with the S/N of the global fringe-fitting gains of the phase calibrator, M\,81*. Such a scaling may not reflect the true natural scatter of the visibilities. Actually, it is common in VLBI to find large scaling factors between the estimated visibility uncertainties and their natural (baseline-based) time scatter, where the latter should better reflect the true visibility uncertainties, assuming that the signal samples of each antenna are independent across the correlator's integration time.

\subsubsection{Global scale of the visibility uncertainties}

There are different ways to estimate the correct uncertainty scaling factor in a VLBI experiment. For instance, \cite{UVMultiFit} scale the parameter uncertainties derived from the post-fit covariance matrix, so that the reduced $\chi^2$ of the model equals its expected value. Such an approach assumes that the fit model is an adequate description of the source brightness distribution and, as such, the residual visibilities are scattered according to their true noise (an assumption that can be checked, for instance, by analyzing the Gaussianity of the residual image).

To perform an homogeneous MCMC exploration of the whole dataset, we applied a similar approach for estimating the global scaling of visibility uncertainties at each epoch. First, we computed the root-mean-squared, rms, of the residual image obtained after a non-parametric deconvolution (in our case, with the CLEAN algorithm, using natural visibility weighting). We then took the rms as a good estimate of the sensitivity of the observations (in units of Jy/beam). Finally, we computed the log-likelihood of a centered point-source in the residuals, and scaled the visibility uncertainties so that the probability density of the flux density of such a point source had a $\sigma$ (in Jy) equal to the image sensitivity estimated from the rms.

\begin{figure}[ht!]
\includegraphics[width=9cm]{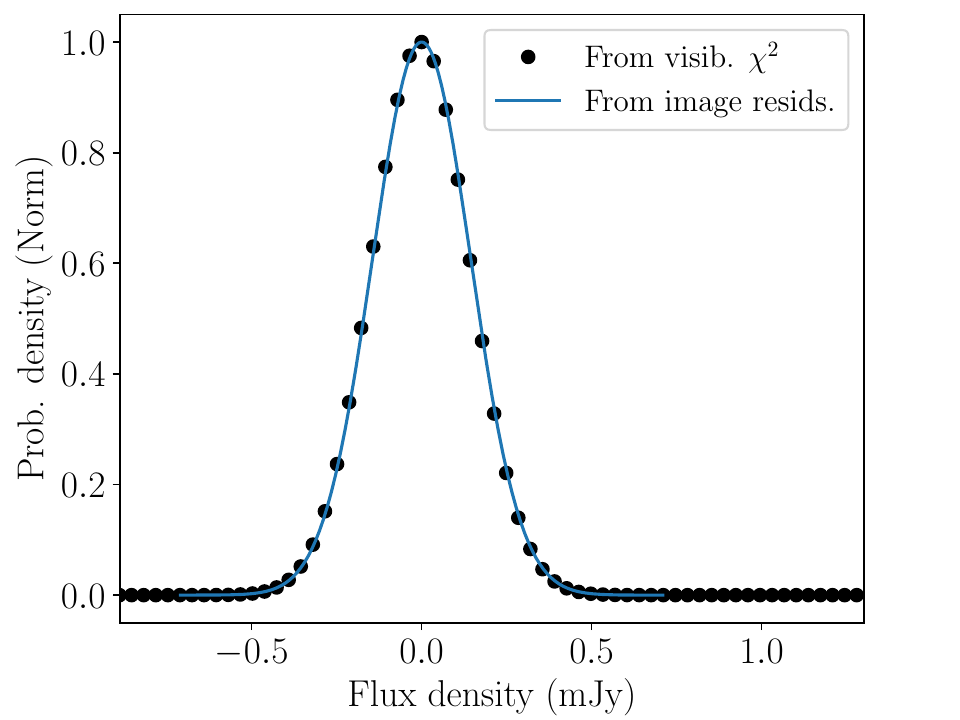}
\caption{Probability density of the flux density of a point source fit to the visibility residuals of day 1639 (C band). Black points are estimates from the likelihood function (after scaling of the visibility uncertainties). The line is the expected distribution, assuming that the image rms (using natural weighting) is equal to the flux-density sensitivity of the observations.}
\label{GaussianFluxFig}
\end{figure}

This strategy is very similar to the standard weight normalization used in any imaging process (i.e., where the sum of all the pixel weights is set to one, so that the image units automatically become Jy/beam). In Fig. \ref{GaussianFluxFig}, we show the probability density of the flux density of a centered point source, fit to the visibility residuals of epoch on day 1639 after explosion, together with the expected Gaussian distribution derived from the image rms. In this figure, the global scaling of visibility uncertainties has already been applied.

\subsubsection{Markov chain samplers}

We sampled the probability densities of the model parameters using Markov chains generated with the Affine Invariant MCMC Ensemble sampler by \cite{Goodman2010}, as implemented in the \texttt{emcee} Python package \citep{EmceeRef}.

We note that the number of free parameters in the model depends on the epoch of observations,  since we included the antenna amplitude factors as model parameters (see Sect. \ref{MethodSec}). 

For each epoch, we generated a set of parallel workers equal to twice the number of parameters. We then updated the chains of each walker in parallel, with a minimum of 500 iterations per walker, until the number of elements per walker is 100 times the estimated (walker-averaged) chain autocorrelation time\footnote{Following the \texttt{emcee} documentation, having more than 50 times the autocorrelation time usually results in a good convergence of the estimators that are derived from the chains}, $\tau$. Once convergence was achieved, we removed the first $2\tau$ elements of the chain, to remove any effects related to the chain initialization.

We notice that if all antennas have free global amplitude gains, and the total flux density of the supernova is also left free, there is a degeneracy in the model that will originate flat (but strongly correlated) posterior distributions for the flux density and antenna gains. To avoid this problem, we set to unity the amplitude correction of the antenna used as reference in the calibration.   

Once the Markov chains are converged, the corresponding estimates of derived quantities (and their uncertainties) are computed from the averages (and standard deviations) of the values obtained from each of the elements in the chain. Since some of the quantities (e.g., relative shell width; see the next section) have a posterior distribution far from Gaussian in some epochs, the standard deviation may not correspond to a confidence interval of 68\% (this interval may even be asymmetric, depending on the skewness of the parent distribution). However, the standard deviations are always directly related to the width of their parent distributions, so they are still good indicators of the parameter uncertainties, even though the Gaussian confidence intervals cannot be directly assigned to them.

\section{Observational results}
\label{ResultSec}

\subsection{Converged Markov chains and biasing effects}

\begin{figure*}[t!]
\centering
\includegraphics[width=0.85\textwidth]{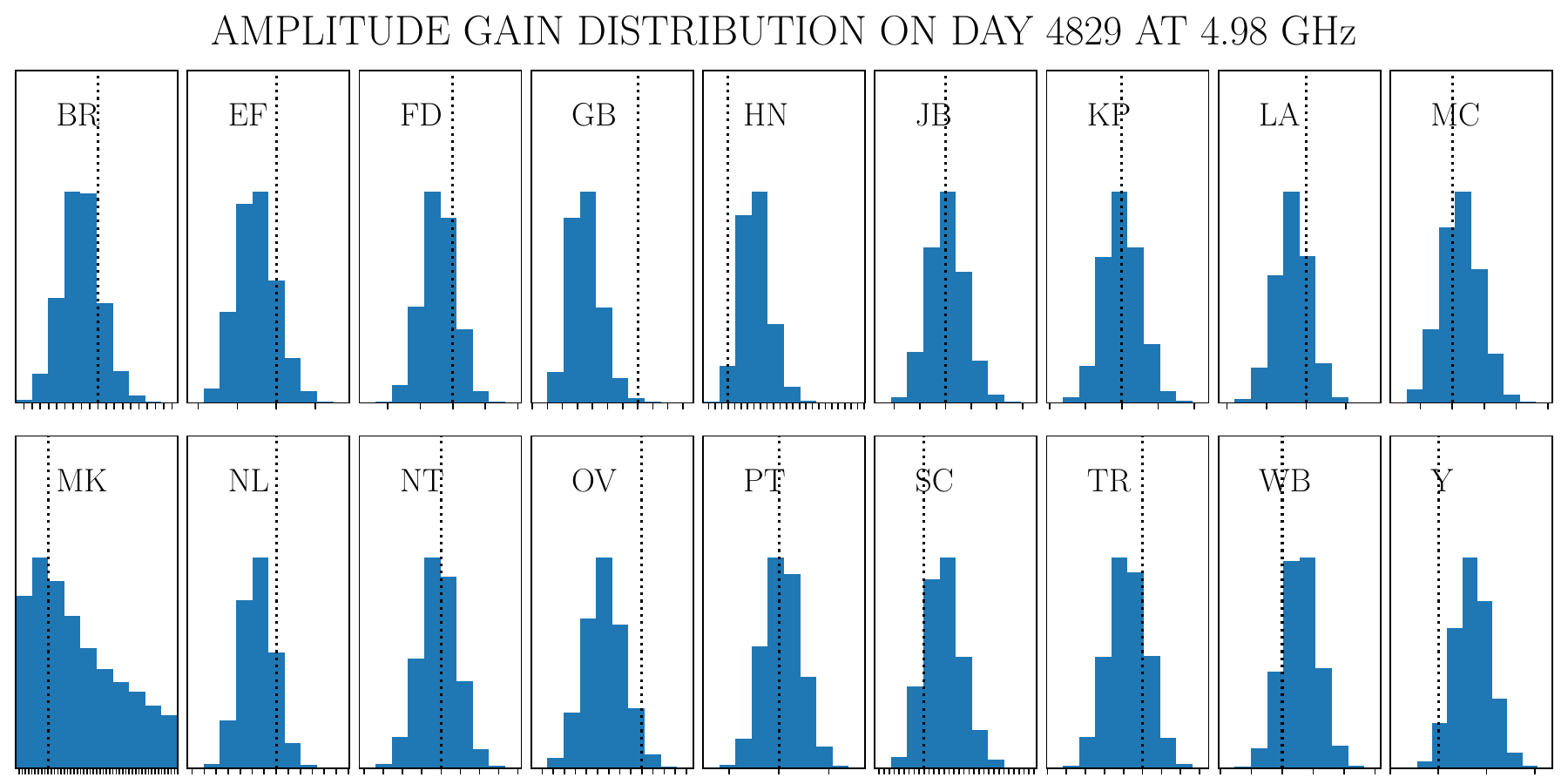}
\caption{Posterior distributions of the antenna amplitude gain factors for the 5\,GHz epoch on day 4829. These are incremental corrections, after applying the $T_{sys}$, gain-elevation curves and global amplitude scaling (based on self-calibration of M\,81*) to the data. Tick marks are separated by intervals of $0.1$. Dotted lines mark the value of a unity gain factor. The histogram for MK has been cut for gains $< 5$, for clarity.}
\label{GainHistoFig}
\end{figure*}

\begin{figure*}[t!]
\centering
\includegraphics[width=0.49\textwidth]{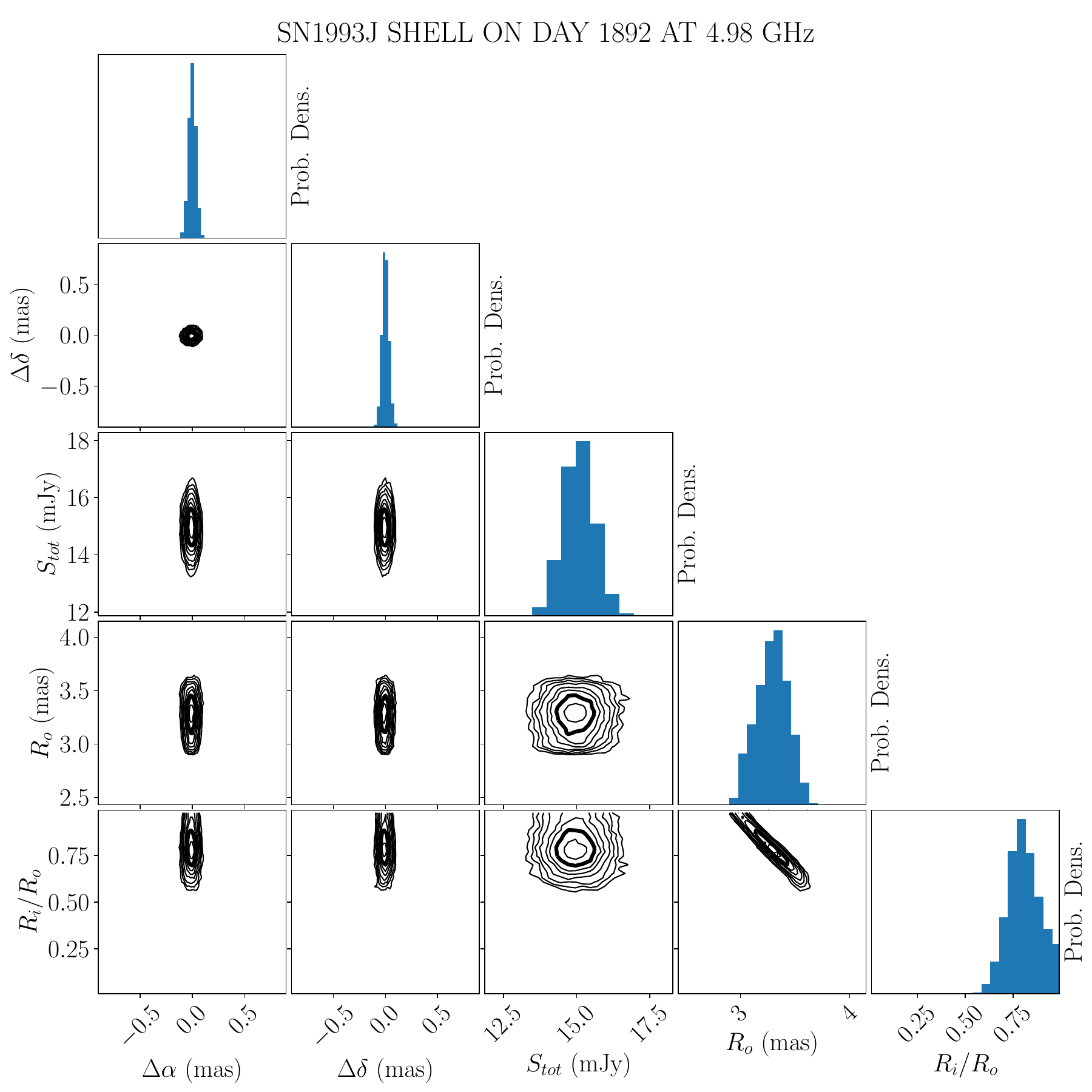}
\includegraphics[width=0.49\textwidth]{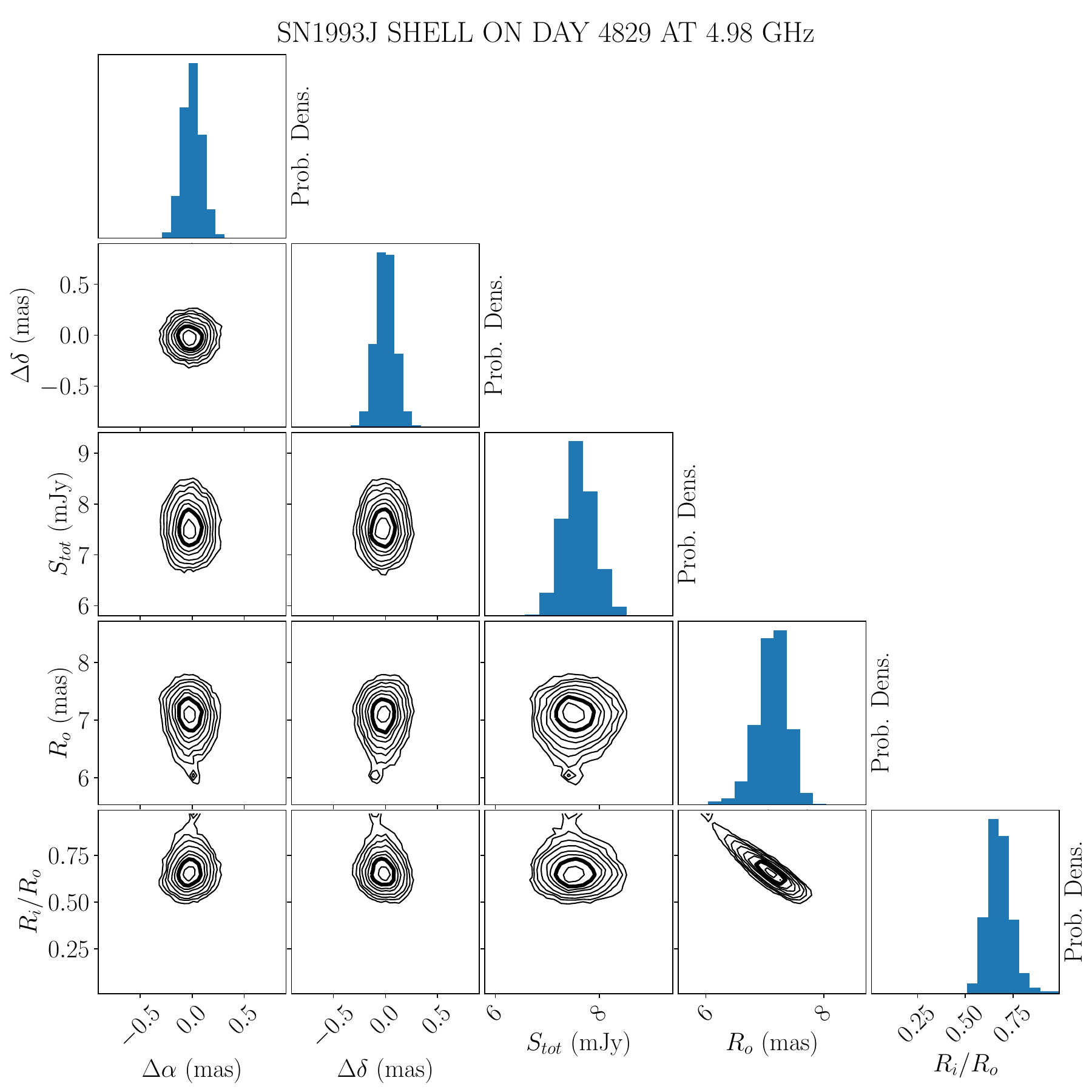}

\caption{Posterior distributions and cross-correlations of shell parameters on day 1892 (left) and on day 4829 at (right), at 5\,GHz. }
\label{SNHistoFigL}
\end{figure*}

As an example of a converged chain, we show in Fig. \ref{GainHistoFig} the final histograms of the antenna amplitude gains for epoch on day 4829 (C band). Notice that, even though most of the antennas have amplitude distributions with peaks close to unity, there are some remarkable exceptions, like Green Bank, Hancock, Owens Valley or the VLA (GB, HN, OV, and Y, respectively). There are also cases (Mauna Kea, MK) where the gain probability distribution is very wide, with amplitude gains even larger than 5. To generate this figure, we have fixed the flux density of the supernova to its average value, mapping the shell flux-density probability distribution into an equivalent amplitude-gain correction for the reference antenna (Los Alamos, LA).

In Fig. \ref{SNHistoFigL}, we show corner plots for the subspace of the supernova shell parameters (i.e., marginalizing the antenna gains) for two epochs (day 1892 at left and day 4829 at right; both at C band). The non-Gaussianity of the shell size and size ratio is clear, especially for the earlier epoch.

\begin{figure}[ht!]
\centering
\includegraphics[width=0.99\columnwidth]{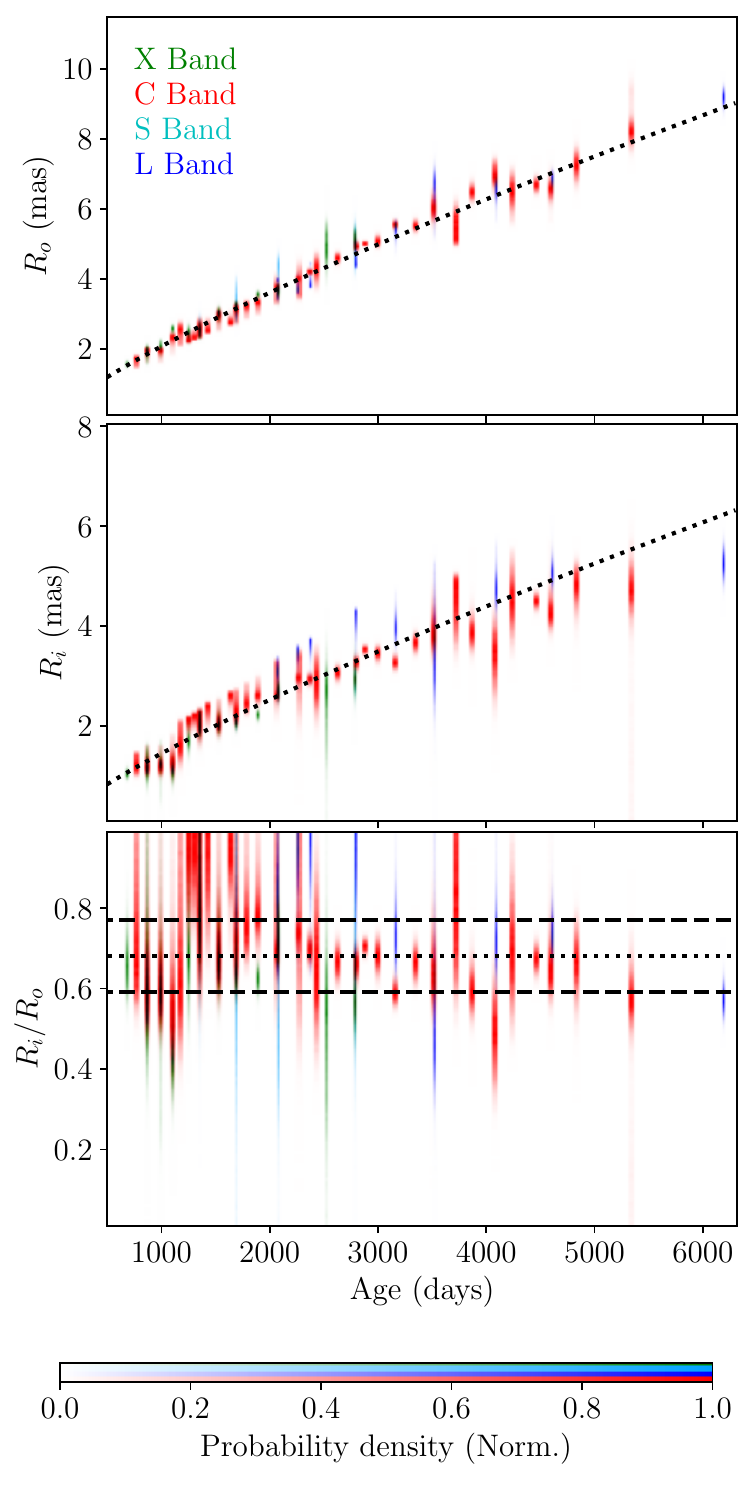}
\caption{Posterior distribution of the time evolution of the supernova shell outer radius (top), inner radius (center) and size ratio (bottom). The dotted lines in the top and center panels show an expansion model with $R_{o} \propto R_{i} \propto t^{m}$ (with $m=0.80$). The dotted (dashed) line in the bottom panel shows the average ($\pm 1\sigma$) size ratio using all the available data (the weighted average of $R_i/R_o$ is $0.69 \pm 0.08$).}
\label{ExpansionFig}
\end{figure}

In Fig. \ref{ExpansionFig}, we show the probability densities of the outer radius, $R_o$, inner radius, $R_i$, and size ratio ($R_i/R_o$, derived directly from the $R_o$ and $R_i$ samples) of SN\,1993J, fit to a spherical shell model with a constant radial intensity profile (i.e., assuming a constant brightness within the shell). All the VLBI epochs, starting from year 1995, are shown in the figure. The probability density distributions are obtained from the converged Markov chains, as is described in Sect. \ref{MethodSec}, which also account for antenna gains. Notice the remarkable asymmetry of the probability distributions of $R_i$ and $R_i/R_o$, for some of the epochs and frequencies.

Even though the prior information used in the MCMC exploration is rather small, there are still data-related and source-dependent effects that may bias the parameter probability distributions shown in Fig. \ref{ExpansionFig}. In particular, we assumed that {\bf 1)} the supernova radio structure is well described by a symmetric spherical shell with no inhomogeneities, {\bf 2)} the shell has a uniform radial brightness distribution, located between radii $R_i$ and $R_o$, and {\bf 3)} the relative visibility uncertainties (i.e., not accounting for the global uncertainty scale discussed in Sect. \ref{MethodSec}) are correct, as determined by the antenna $T_{sys}$ and the S/N of the fringes. Regarding the last point, we may still need to apply an extra radial weighting to the visibilities (which affects the relative visibility weights), in order to avoid (or minimize) running biases, as was discussed in \cite{Marcaide2009} and \cite{MartiVidal2011a} (see also the next subsection). However, we notice that the parameters for such a visibility weighting are arbitrary.

In the following subsections, we analyze how departures from any of the assumptions summarized in the previous paragraph may affect the estimates of the shell parameters shown in Fig. \ref{ExpansionFig}.

\subsection{Effect of radial visibility weighting}

As the supernova shell expands, the Fourier transform of its brightness distribution becomes more compact in the UV space. Therefore, the signal encoded in the longer baselines becomes weaker (in units of the total flux density), whereas the higher visibility amplitudes \citep[corresponding to the main lobe and first sidelobes of the Hankel transform of the shell; see][]{Marcaide2009} get packed in shorter 
baselines. 

One way to minimize possible model-fitting biases related to this effect is to apply a radial weighting to the visibilities that scales with the source size. In particular, a Gaussian UV taper can be applied, so that the longer baselines are progressively down-weighted in the fit as the shell expands. This strategy was used in \cite{Marcaide2009} and \cite{MartiVidal2011a} to remove possible running biases in the shell size estimates. However, we notice that the particular selection of the UV-taper size (to be scaled to the inverse of the shell size) is rather arbitrary. In any case, rescaling the UV taper affects the shell-size estimates, but in a way that does not affect the expansion index, $m$ \citep[being $R_o \propto t^m$, as was discussed in ][]{Marcaide2009}.

However, that reasoning is only correct if the observations are not strongly limited by noise. The use of UV tapers may have an additional impact if the observations are severely S/N limited. The most sensitive baselines in VLBI are usually the shorter ones, so that using UV tapers to weight the long baselines down could be understood as the application of adaptive filters, to enhance the S/N of the detections. There is thus the possibility that the use of UV tapers will have extra effects on the observations with lower S/N (i.e., the latest epochs of the campaign, when the supernova brightness was the lowest). In such a case, the Markov chains will already take into account all the effects of UV tapering.

\begin{figure*}[ht!]
\centering
\includegraphics[width=0.45\textwidth]{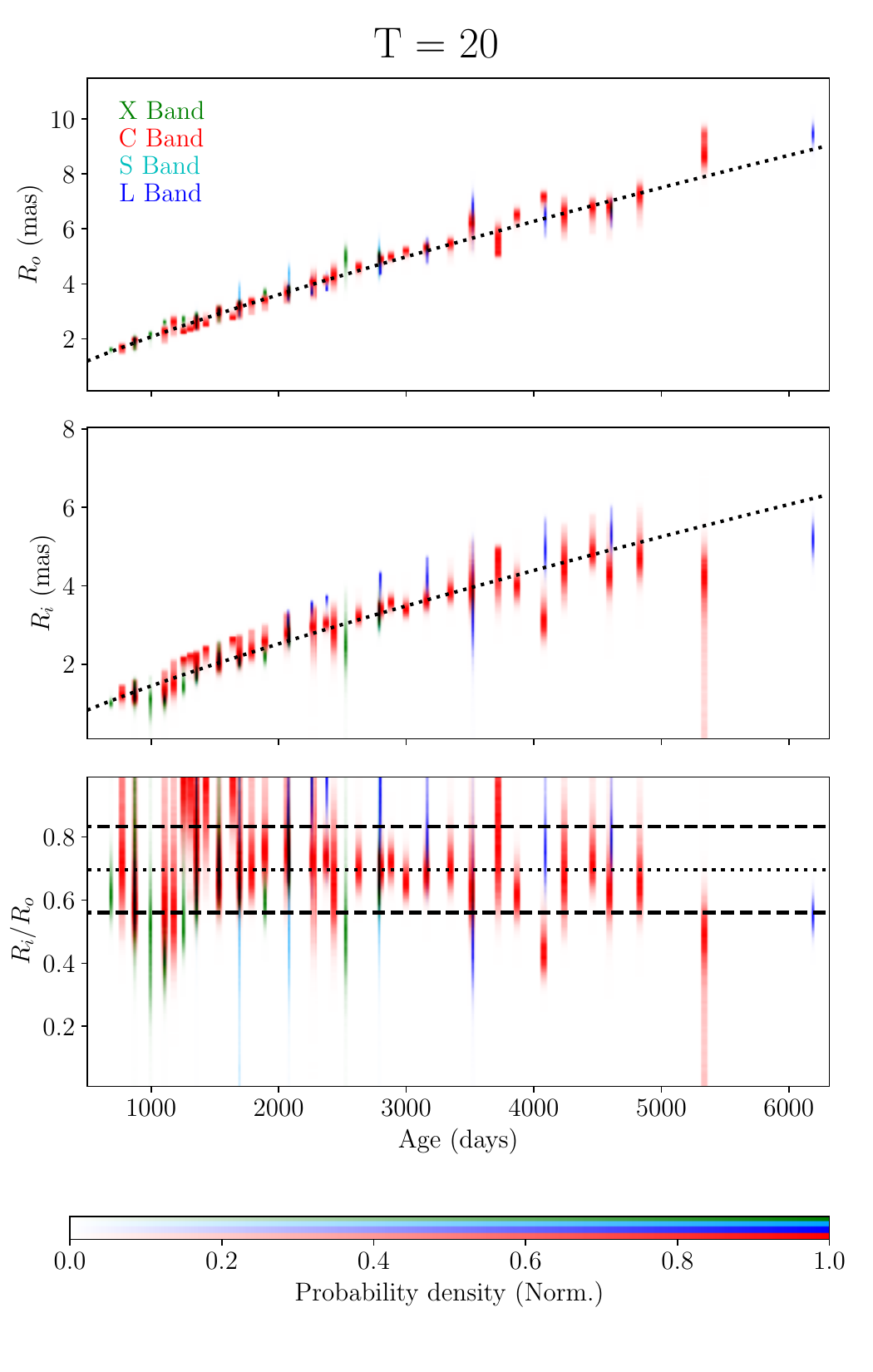}
\includegraphics[width=0.45\textwidth]{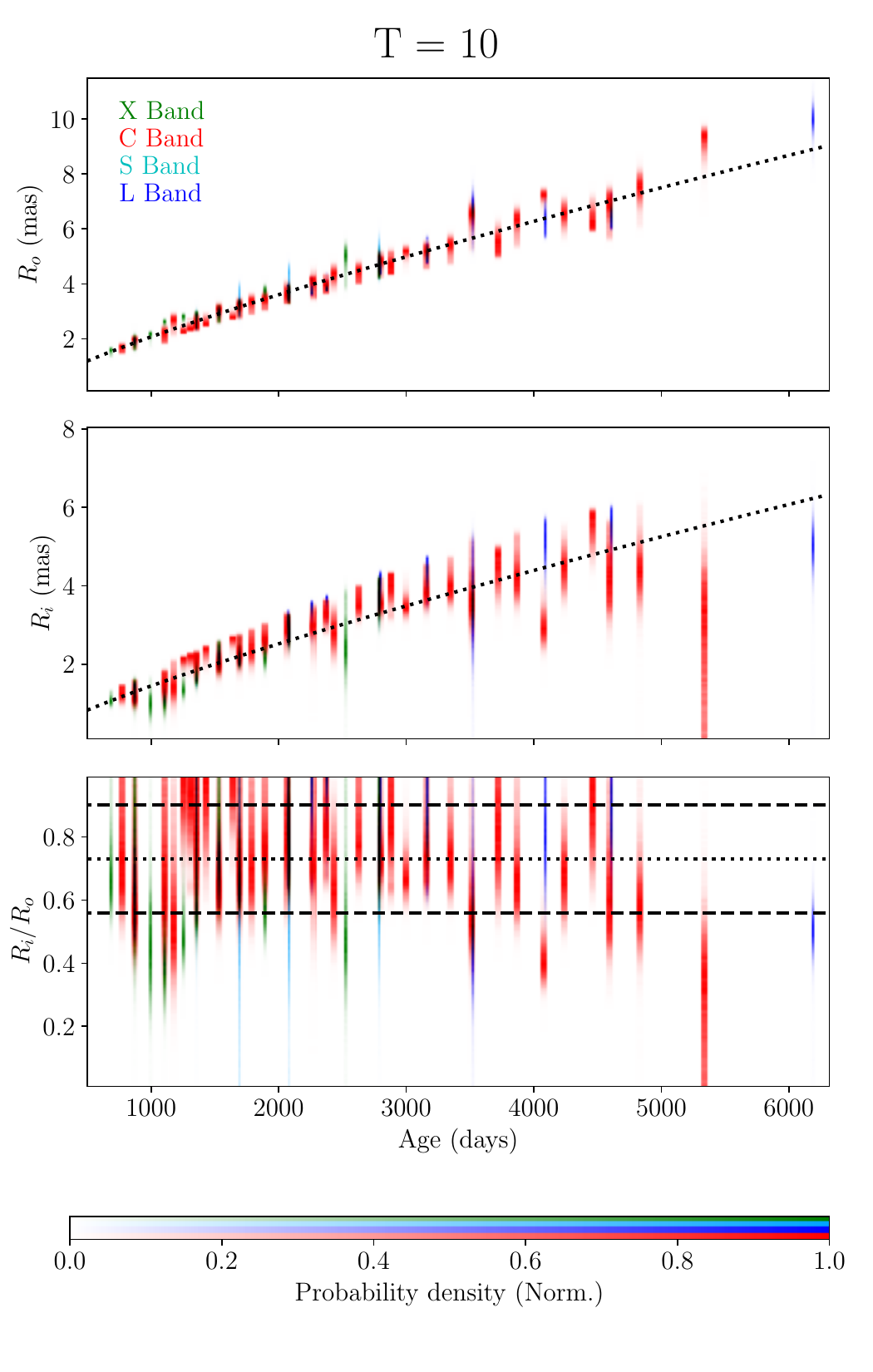}
\caption{Same as Fig. \ref{ExpansionFig}, but using a UV taper with HWHM given by Eq. \ref{taperEq}, with $T = 20$ (left) and $T=10$ (right). The statistics for $R_i/R_o$ are $0.70\pm 0.15$ and $0.73 \pm 0.18$ for $T = 10$ and $T = 20$, respectively.}
\label{ExpansionUVTaperFig}
\end{figure*}

We have thus repeated the run of Markov chains described in Sect. \ref{MethodSec},  adding additional weighting factors to the visibilities. The UV tapers used in this analysis are Gaussians in Fourier space, with half width at half maximum, $\Theta$, given by

\begin{equation}
    \Theta \, \left[ M\lambda \right] = \frac{T}{R_o \, \left[mas\right]},
    \label{taperEq}
\end{equation}

\noindent where $R_o$ is the estimated outer radius of the supernova and $T$ is an arbitrary constant. For the UV tapers used in \cite{Marcaide2009} and \cite{MartiVidal2011a}, we have $T \sim 30$. In this paper, we have used different values of $T$, to test how different choices of UV tapers may affect the shell parameters. In Fig. \ref{ExpansionUVTaperFig}, we show the results for $T= 20$ (left) and $T = 10$ (right).

A direct comparison between Fig. \ref{ExpansionUVTaperFig} and \ref{ExpansionFig} indicates that using narrower UV tapers (i.e., lower values of $T$, which result in a stronger down-weighting of long baselines) increases the widths of the probability density distributions of all quantities, $R_o$, $R_i$ and $R_i/R_o$. This is especially true for the latest epochs, which are more severly affected by noise. Actually, the model on day 5334 at C band, using the narrower UV taper, has a shell width with a high probability density even for values of $R_i/R_o$ close to 0 (which would correspond to a uniform filled sphere). As a consequence, the probable values of $R_o$ also increase (since the shell width and size are correlated parameters in the model). Higher probability densities for (slightly) larger shell sizes make the results depart from the power law expansion model (top panels of Fig. \ref{ExpansionUVTaperFig}) and give a false hint of reacceleration of the shell (compare the two last epochs in the top panel of Fig. \ref{ExpansionFig} to those of Fig. \ref{ExpansionUVTaperFig}). In Sect. \ref{DiscussSec}, we discuss about this model-driven reacceleration effect and its interpretation in the literature \citep{bie11}.

\subsection{Effect from shell inhomogeneities}

The images of \SN93J\ do not have perfect azimuthal symmetry. \cite{Bartel2002}, \cite{Marcaide2009} and \cite{MartiVidal2011a} quantified in different ways the degree of circularity and the level of inhomogeneities, across the shell expansion history, and all authors found clear signs of asymmetries and inhomogeneities, albeit at the level of only a few percent.  Actually, \cite{InstrumentalShell} simulated realistic VLBI synthetic datasets of SN1993J and concluded that the small observed inhomogeneities might actually be explained by the limited VLBI image fidelity, being the true shell of \SN93J more symmetric and homogeneous than what is inferred from the VLBI images. In any case, and to a good first approximation, we can assume that the images of \SN93J are circularly symmetric. 

However, even for such a low level of inhomogeneities, there may still be small biases in the estimated shell sizes and/or widths, which are based on the assumption of a perfect spherical symmetry. We estimated the level of bias due to the presence of inhomogeneities in the shell, using synthetic data, as is explained in the following lines.

We generated a synthetic VLBI dataset, corresponding to a supernova shell with a perfect spherical symmetry, and computed the corresponding Markov chains. We then introduced realistic inhomogeneities to the data, based on an actual \SN93J image, and computed new Markov chains, so that the comparison between the posterior distributions from both chains are a good indication of the contamination that the inhomogeneities may imprint into the parameter probability distributions.

\begin{figure*}[ht!]
\includegraphics[width=0.95\textwidth]{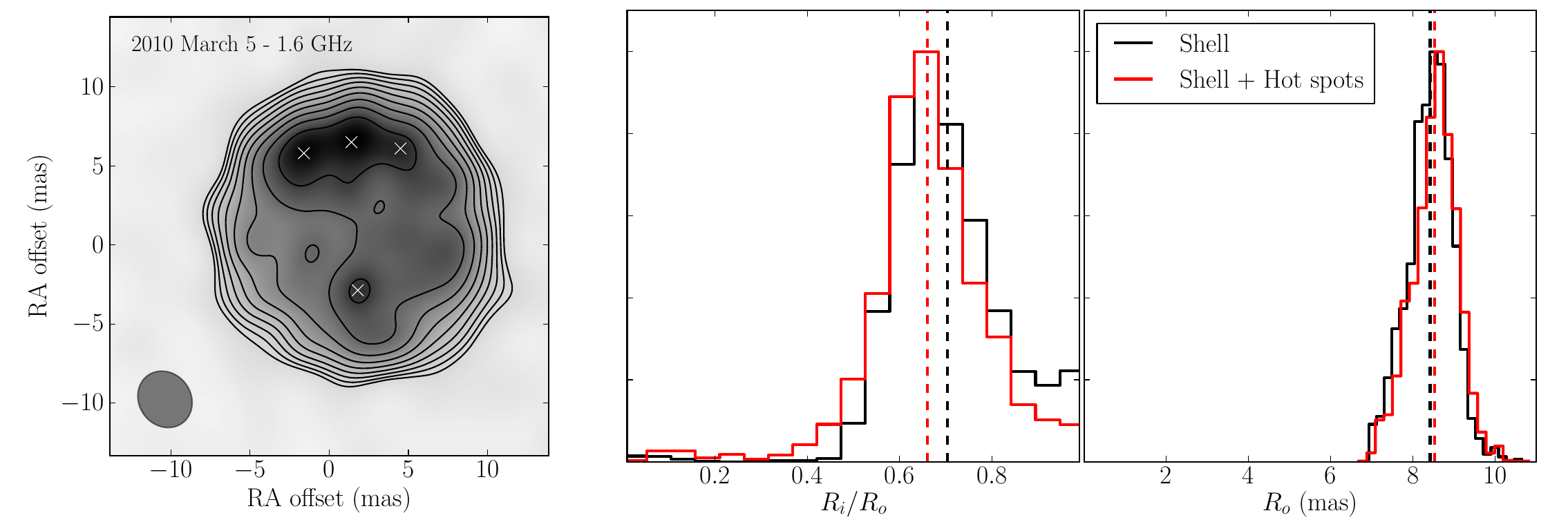}
\caption{%\LEt{***The first sentence of each figure caption or legend should be a descriptive title of the figure as a whole, written in a telegraphic style with the initial article (the, a, an) omitted. The text that follows should be written in complete sentences. If individual panels are described, this is repeated (another telegraphic description followed by full sentences).}
Image used to study the effect of shell inhomogeneities on the expansion curve. Left, CLEAN image of SN\,1993J at L band on March 5, 2010. The 10 contours are set logarithmically, from the image peak (0.12\,mJy/beam) to 10\% of the peak. The FWHM of the convolving Gaussian beam is shown at the bottom left corner ($3.72\times3.28$\,mas, with a position angle of $-36$ degrees). The best-fit shell model for this epoch is used to generate synthetic observations, from which the posterior shell parameter distributions are determined (black histograms; center and right panels). Then, the most prominent hot spots in the image (white crosses in the left panel) are added to the synthetic data, to determine new parameter distributions (red histograms). The dashed lines at the center and right panels mark the parameter averages for each case.}
\label{HotSpotFig}
\end{figure*}

The synthetic data were generated based on epoch from 5 March 2010 at 1.4\,GHz. Since this is the last VLBI epoch of \SN93J, the brightness peak of the shell is minimum, so that the effects of inhomogeneities and noise are likely a conservative representation of what is present in the rest of observations. The shell parameters used in the simulation correspond to those of the maximum probability density from the Markov chains obtained from the real data. In addition to this, we also added realistic noise and antenna-gain corruptions to the simulations, based on the results obtained from the Markov chains of the real data. The inhomogeneities added to the synthetic spherical shell were taken from the CLEAN image of that same epoch, and are shown in Fig. \ref{HotSpotFig} (left panel). The histograms corresponding to the shell size, $R_o$, and size ratio, $R_i/Ro$, are shown in the same figure (center and right panels), both for the case of a perfect shell (black lines) and of a shell with inhomogeneities (red lines). 

The main conclusion from these results is that, even though the effect of inhomogeneities is negligible for the estimate of $R_o$, it seems to have a mild effect on the estimate of $R_i/R_o$ (adding inhomogeneities decreases the ratio estimate by $\sim 7$\%). In any case, the amount of this bias is of the order of the uncertainties that can be inferred from the probability-density distributions shown in Fig. \ref{ExpansionFig} (lower panel). Hence, it is not likely that the presence of inhomogeneities will strongly affect the conclusions drawn from our estimates of $R_o$ and $R_i/R_o$ beyond the statistical uncertainties.

\subsection{Effect of the radial intensity profile}
\label{RadialProfileSec}

All shell models considered in this paper until now, as well as in all previous VLBI works on \SN93J \citep[e.g., ][]{Bartel2002, Marcaide2009, MartiVidal2011a, MartiVidal2011b}, have assumed a uniform brightness distribution between $R_i$ and $R_o$. This assumption implies that the electron density and magnetic-field energy density are distributed homogeneously and isotropically within the shell. 
We have analyzed the effects of departures of the shell radial brightness profile from uniformity, by adding new shell parameters that account for (spherically symmetric) radial variations in the brightness. In particular, we have parameterized the radial intensity profile using two different models. 

First, a linear model with the radius, $r$, and a slope, $R_{sl}$, which is bounded between $R_{sl} = -1$ (corresponding to a maximum intensity at $R_i$ and a zero intensity at $R_o$) and $R_{sl} = +1$ (zero intensity at $R_i$ and maximum intensity at $R_o$); a value of $R_{sl} = 0$ corresponds to a uniform brightness distribution. Second, a Gaussian radial intensity profile, bounded between $R_i$ and $R_o$, with a peak location, $\rho_{peak}$ (relative to $R_o$), and a fixed width equal to half of the shell width.%\LEt{***Single-sentence paragraphs should be avoided.}  
The actual equation used for the linear radial intensity profile, $I(r)$, is

\begin{equation}
    I(r) = \frac{1}{2} + \frac{R_{sl}}{R_o-R_i}\left(r - \frac{R_o+R_i}{2}\right) ~~\mathrm{for}~~r \in [R_i,R_o],
\end{equation}

\noindent whereas for the Gaussian profile, it is

\begin{equation}
    I(r) = \exp{\left[ \frac{-2(r - R_o\rho_{peak})^2}{(1-R_i/R_o)^2} \right]} ~~\mathrm{for}~~r \in [R_i,R_o].
\end{equation}

\begin{figure*}[ht!]
\sidecaption
\includegraphics[width=12cm]{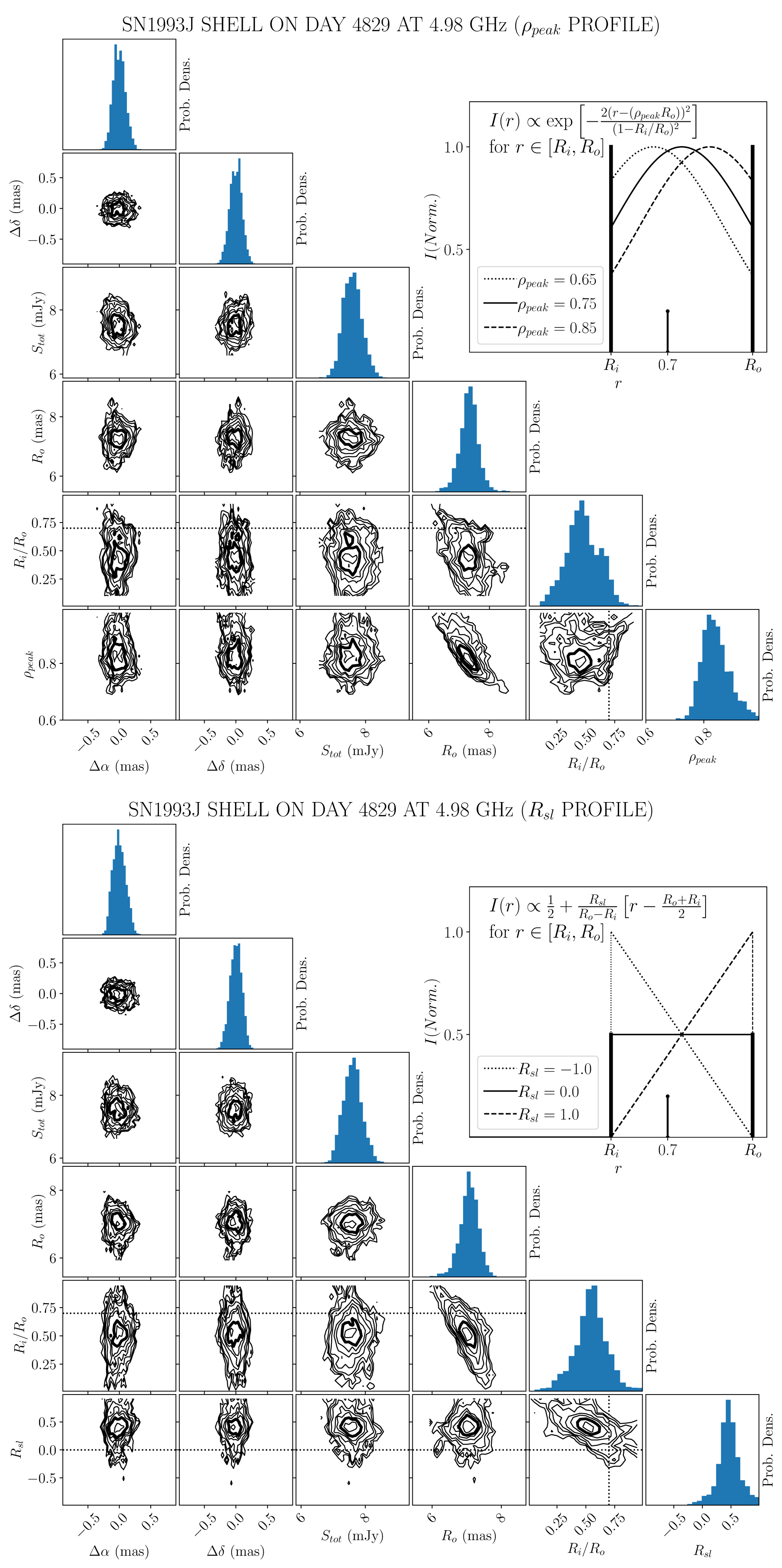}
\caption{Corner plots of the shell parameters for epoch on day 4829 at 5\,GHz for a Gaussian radial intensity profile (top) and a linear radial profile (bottom). The value $R_i/R_o=0.7$ is marked as dotted lines. In the upper-right part of each corner plot, examples of radial intensity profiles are shown assuming $R_i/R_0 = 0.5$.}
\label{SNHistoProfileFig}
\end{figure*}

\noindent
We show example plots of these radial intensity distributions in Fig. \ref{SNHistoProfileFig}. The values $R_{sl}=0$ (i.e., uniform shell) and $\rho_{peak} = (1+R_i/R_o)/2$ (i.e., centered peak brightness), are plotted as continuum lines, for clarity (other distributions are shown as dotted and dashed lines). A value of $R_i/R_o = 0.5$ has been used. This  
 produces new Markov chains with some interesting properties. We show example corner plots of converged chains in Fig. \ref{SNHistoProfileFig} for the model with $R_{sl}$ (top) and $\rho_{peak}$ (bottom), corresponding to day 4829 after explosion, at 5\,GHz.

\begin{figure*}[ht!]
\centering
\includegraphics[width=0.49\textwidth]{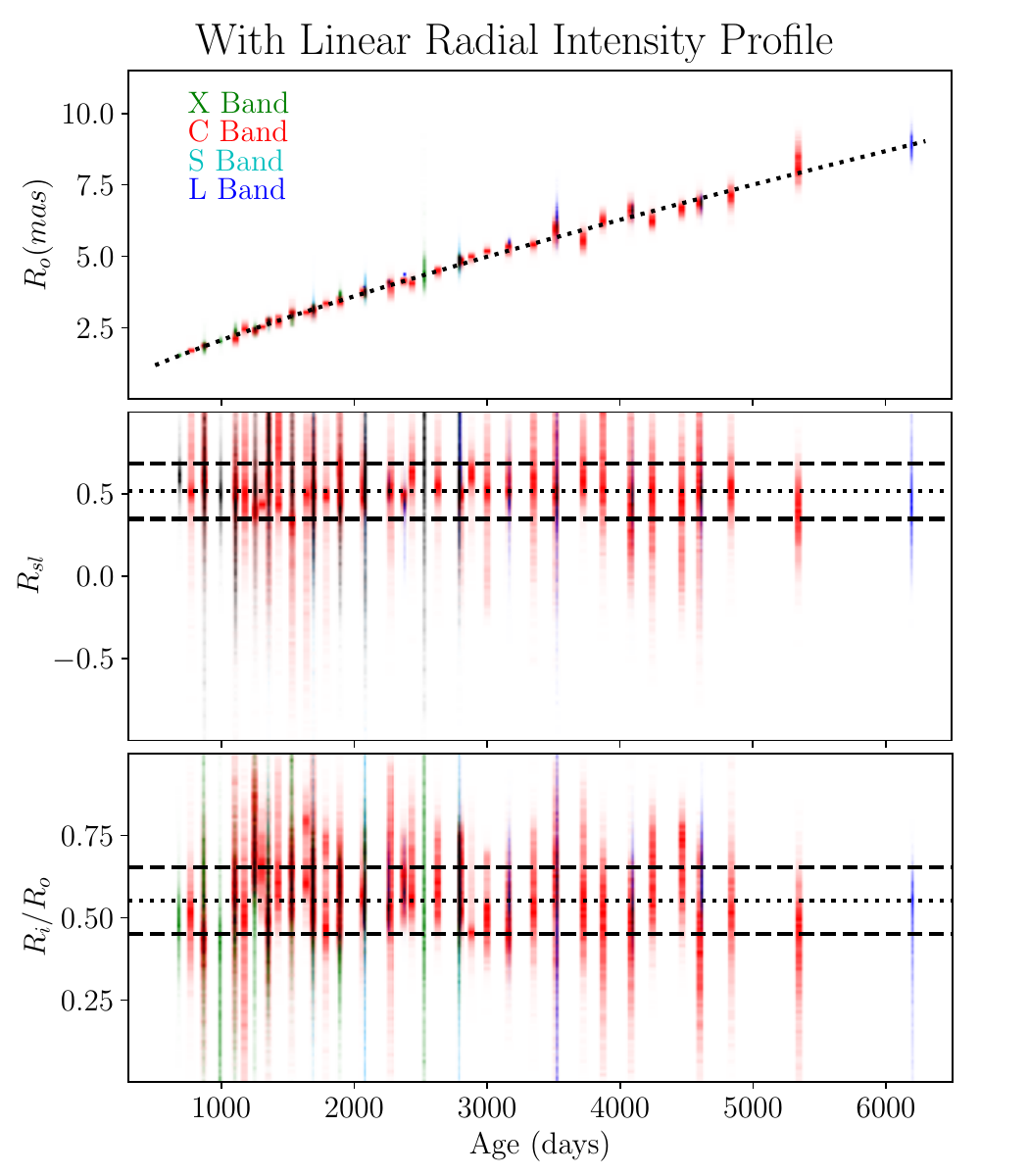}
\includegraphics[width=0.49\textwidth]{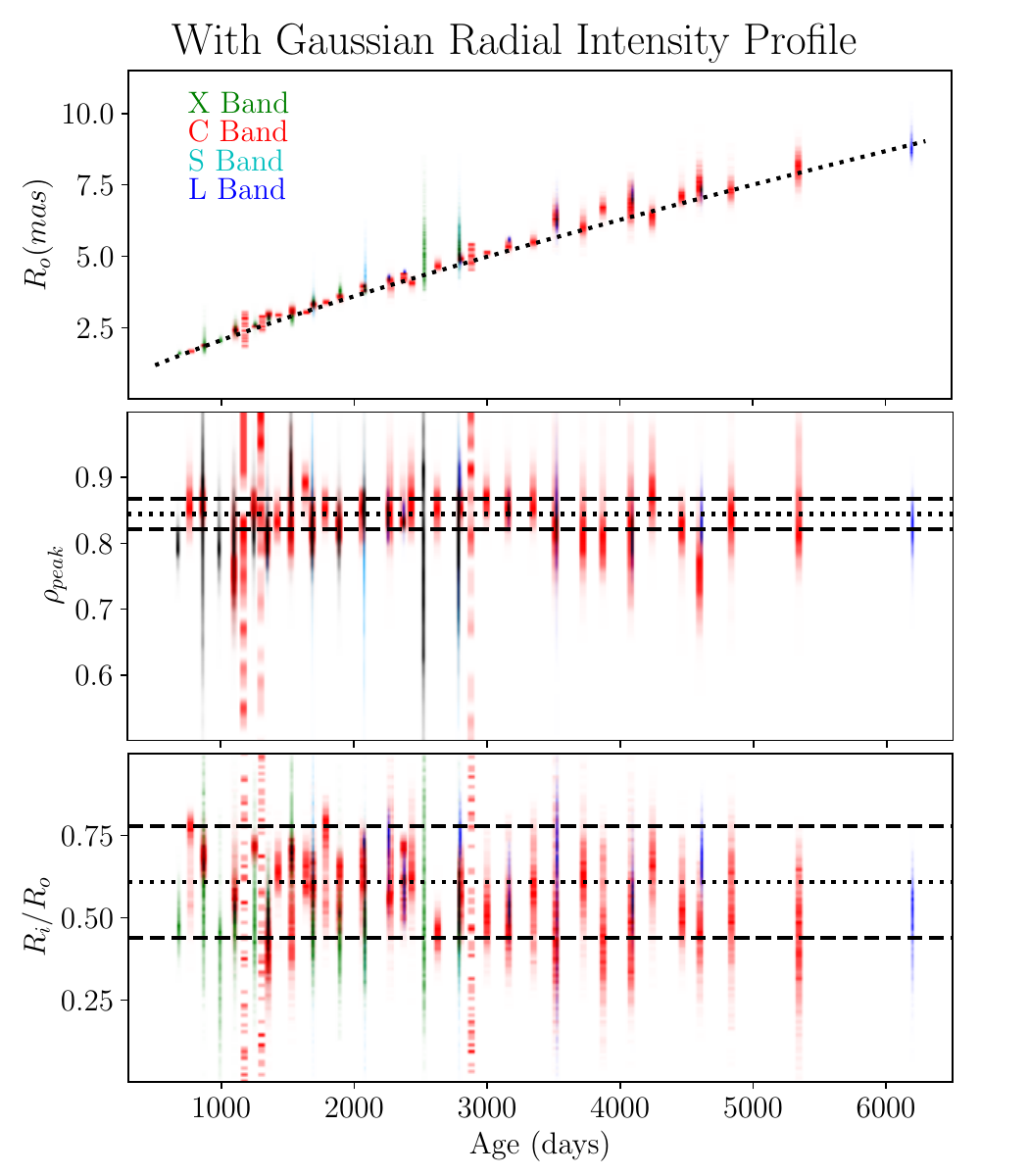}
\caption{Posterior shell parameter distributions for the model with a linear (left) and Gaussian (right) radial intensity profile. Top panels, shell size; center panels, radial intensity parameters (slope, $R_{sl}$, at left; peak position, $\rho_{peak}$, at right); bottom panels, $R_i/R_o$. The color codes are the same as in the previous figures. Dotted lines in the top panels show the power law expansion model $R_o \propto t^m$ (with $m=0.8$). Dotted (dashed) lines in the center and bottom panels mark the average ($\pm 1\sigma$) values for the whole campaign.}
\label{SlopeFig}
\end{figure*}

\paragraph{Linear intensity profile.} A remarkable result to notice  (see, for instance, the correlation plot between $R_{sl}$ and $R_i/R_o$ in Fig. \ref{SNHistoProfileFig}, bottom) is that the probability density increases notably toward positive values of $R_{sl}$, which is strong evidence of a higher shell emissivity closer to the forward shock. Another 
result is that, even though the best $R_i/R_o$ estimate for a uniform shell ($R_{sl}=0$) is around $\sim 0.7$ (see the cross between the dashed lines and the 2D histogram in Fig. \ref{SNHistoProfileFig}, top), the preference of the model toward positive $R_{sl}$ also translates into lower values for $R_i/R_o$ (i.e., wider shell widths, same panel in the figure). This is to be expected, since modeling a radially increasing intensity under the assumption of a uniform distribution would artificially increase the value of the inner boundary. These two conclusions can actually be generally applied to the rest of observations in the campaign. In Fig. \ref{SlopeFig} (left, center panel), we show the posterior probability-density distributions of $R_{sl}$ for all epochs, following the same color codes as in the previous figures. In all cases, with no exception, the $R_{sl}$ distributions take positive values, typically in the range between 0.35 and 0.65, which correspond to intensities at $R_i$ equal to $20-30$\% of those at $R_o$. Therefore, all the \SN93J data favor a higher intensity closer to the forward shock and decreasing toward the inner regions of the shell.
The statistical preference over a uniform shell is also robust: the increase in probability density for the chain samples of the model with $R_{sl}$ is typically several factors (or a factor of several tens) larger than those of the uniform shell. %\LEt{***Please check that your intended meaning has been preserved.} 
A simple $\chi^2$ comparison test favors the model with the extra parameter, $R_{sl}$, with a significance that, in some epochs, reaches values $\sigma \gg 5$.

\paragraph{Gaussian intensity profile.}  
Although the linear intensity profile indicates that the emission is dominated by the outer part of the shell, this does not necessarily imply that it originates close to the forward shock. The location of the peak in the radial brightness distribution can be estimated with the use of the Gaussian profile.
In Fig. \ref{SlopeFig} (right, center panel), we show the distributions of $\rho_{peak}$ for all epochs\footnote{We notice that convergence of the chains could not be properly achieved in three of the epochs, which are clearly seen in the figure.}. The weighted average of $\rho_{peak}$ for all epochs before day 3500 is $\rho_{peak}^{early} = (0.850 \pm 0.003)$. For all the later epochs, the average is $\rho_{peak}^{late} = (0.823 \pm 0.010)$, which is $\sim 2\,\sigma$ lower (i.e., mild evidence of a decrease in the emissivity around the forward shock). One may also note that the distribution of $R_i/R_o$-values is roughly the same as for the linear intensity profile. The fact that two qualitatively different parameterizations of the radial intensity distribution give similar overall shell properties, strengthens the reality of the non-uniform shell structure.

\section{Discussion}
\label{DiscussSec}

\subsection{Comparison with previous results}

\subsubsection{Supernova expansion}

The time evolution of $R_o$ is well described by a power law with time, $R_o \propto t^m$, with an expansion index of $m=0.802\pm0.010$ (fit to the data shown in Fig. \ref{ExpansionFig}, which corresponds to the case of a shell with a uniform radial intensity profile). The same expansion index produces satisfactory fits for data with different choices of UV tapers (Fig. \ref{ExpansionUVTaperFig}) and for a model with varying radial intensity profiles (Fig. \ref{SlopeFig}).

In previous publications \citep[e.g.,][]{Bartel2002, Marcaide2009, MartiVidal2011a, MartiVidal2011b}, different expansion indices covering different time ranges were reported, in order to properly match the late evolution with the very early observations ($m$ being closer to unity at early epochs, when the expansion is almost free, and decreasing at later epochs). For instance, \cite{Marcaide2009} found that $m=0.85$ for $t<1500$ days after explosion, followed by $m=0.79$ for later times. In this publication, however, we prefer to use a much simpler model, with one single expansion index, since we are especially interested in the late supernova evolution and, in any case, the fit with $m=0.802$ already produces satisfactory results for the whole time range covered.

\subsubsection{Shell width}

The values of $R_i/R_o$ (or, equivalently, the shell width, $\rho = (R_o-R_i)/R_o$) reported in the literature are quite different, depending on the authors and the methods used to estimate them. For instance, \cite{MarcaideNat} estimated $\rho \sim 0.3$, which was later confirmed by \cite{Marcaide2009}, \cite{MartiVidal2011a} and \cite{MartiVidal2011b} with a final estimate of $\rho = 0.31 \pm 0.02$. On the other hand, \cite{Bartel2000} reported a value as low as $\rho = 0.205\pm0.015$, which was later revised by \cite{Bietenholz2003} to $\rho = 0.25\pm0.03$. 

In this work, and based on the complete parameter exploration of our Markov chains, we get an average (over the whole campaign) value of $\rho = 0.31 \pm 0.08$ for a uniform radial intensity distribution. We notice that the large uncertainty associated with the widths of the posterior size-ratio distributions reflects the effect of noise and antenna gains. It is also worth noticing that the probability density distributions of $\rho$ are, for many epochs, not symmetric (i.e., Gaussian-like) and show clear skewness toward either lower or higher values, depending on the epoch (see, for instance, the histograms in Fig. \ref{SNHistoFigL}). Having skewed distributions for $\rho$ may result in small biases for the average (and error estimation) among all epochs.

At late epochs, there is also a hint of a departure of the shell from a self-similar expansion. This can be seen in Fig. \ref{ExpansionFig} (center panel), where there is very little change in $R_i$ from day $\sim$4000 (the values fall below the prediction of the expansion model), while the growth in $R_o$ persists during the whole campaign (top panel). This behavior may be indicative of a slight widening of the shell at the latest epochs. Such an effect was also seen 
in \cite{bie11} and will be further analyzed in Sect. \ref{DiscussSec2}. 

Actually, the weighted average of $\rho$ for all epochs before day 3500 is $\rho_{early} = 0.312 \pm 0.006$, whereas the average for all the later epochs is $\rho_{late} = 0.356 \pm 0.014$. There is thus mild evidence of a widening in the shell at late epochs with a confidence of 2.3$\sigma$. We can also test the shell widening hypothesis by proposing an expansion model for $R_i$ with a changing power index. We have fit the values of $R_i$ with a simple model with two expansion indices, $m_1$ and $m_2$, separated by a break time, $t_{br}$, similar to the model used in \cite{Marcaide2009}, although, in that case, the fit quantity was the outer radius, $R_o$.

\begin{figure*}[ht!]
\centering
    \includegraphics[width=6cm]{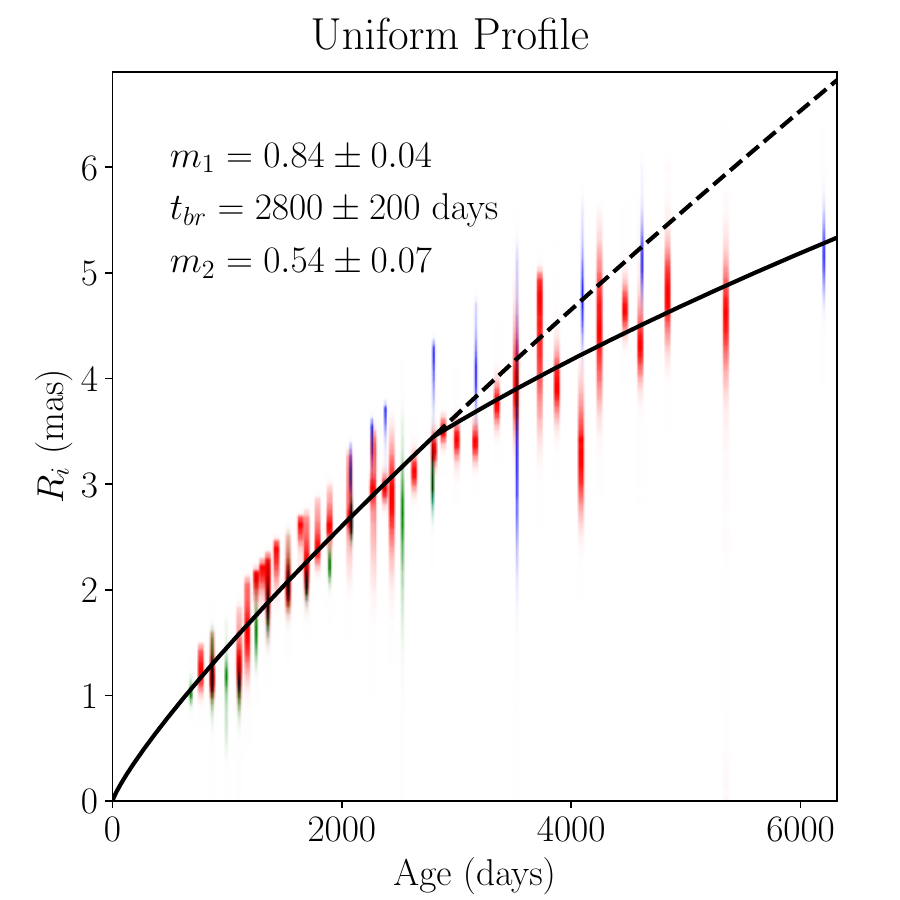}
    \includegraphics[width=6cm]{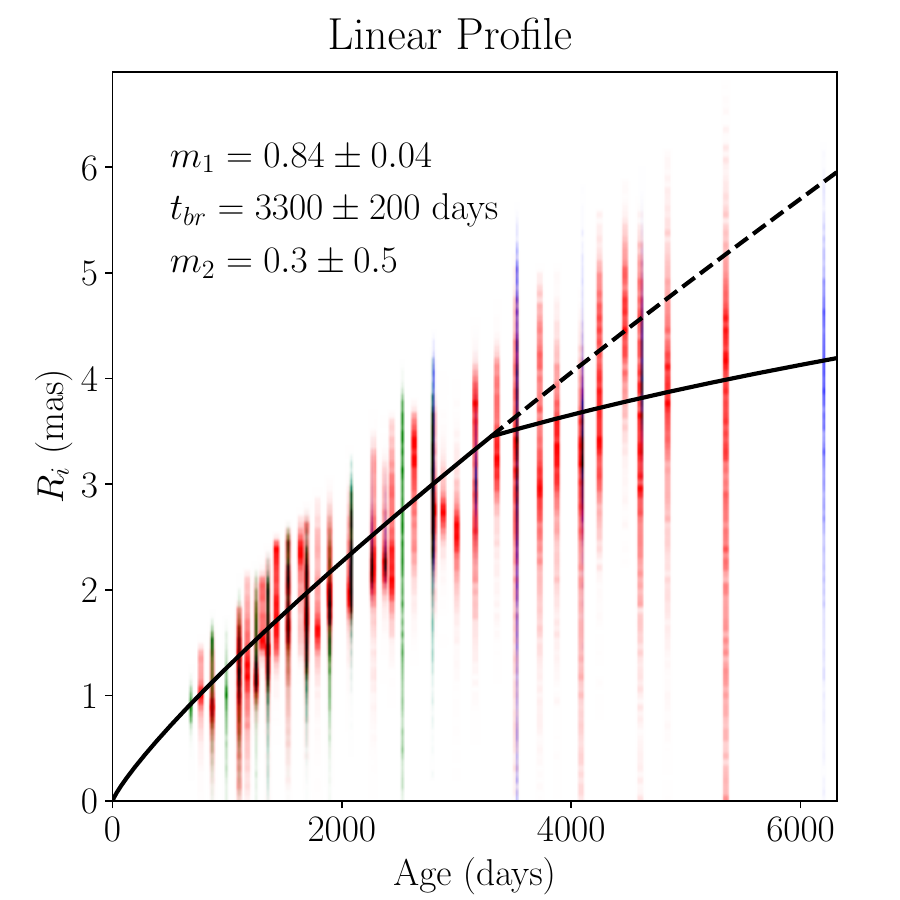}
    \includegraphics[width=6cm]{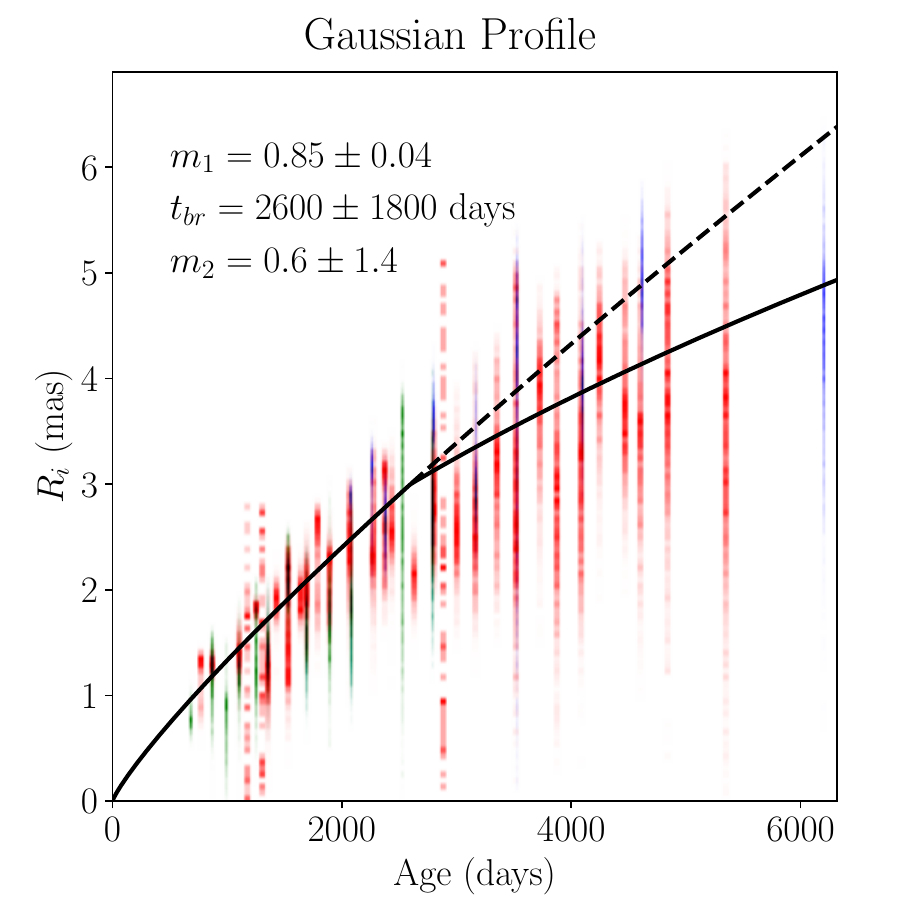}
    \caption{Expansion of the inner shell radius, $R_i$, using the model of a uniform spherical shell (left) a linear radial profile (center) and a Gaussian radial profile (right). The color codes are the same as in the previous figures. The data are fit to an expansion model (thick lines) with two power indices, $m_1$ and $m_2$, separated by a time break, $t_{br}$. The extrapolation of the expansion with $m_1$ is shown as dashed lines.}
    \label{fig:RiExpansion}
\end{figure*}

In Fig. \ref{fig:RiExpansion}, we show the expansion curves of $R_i$ estimated with different radial intensity profiles (uniform, linear and Gaussian). The posterior distributions of $R_i$ for the models with nonuniform brightness profiles are notably wider than those for the uniform shell model. This is likely related to the extra degrees of freedom in the nonuniform models, whose posteriors, correlated with $R_i$, will result in a wider posterior for $R_i$. This may also be indicative of some bias in the uniform-model posteriors of $R_i$ if the true brightness distribution is different from uniform. In such a case, the $R_i$ posteriors corresponding to the linear and Gaussian profiles might be less biased (although less precise) than in the uniform model. 

The expansion curves of $R_i$ in Fig. \ref{fig:RiExpansion} are fit to a model with two power indices. In all cases, there are indications of a break time, $t_{br}$, around $2500-3500$\,days after explosion, with the power index decreasing from $m_1 \sim 0.84$ to only $m_2 \sim 0.5$. This is an indication of a widening in the shell width at late epochs (i.e., an increased deceleration of the reverse shock), a conclusion that is independent of the radial profile considered. It may be noted, though, that the widening of the shell is more rapid for the nonuniform brightness profiles than for the uniform one; that is, at a given time, the shell width is smallest for the uniform model. This widening appears clearly if we compare the late values of $R_i$ to the model predictions with one single power index (i.e., the dashed lines in Fig. \ref{fig:RiExpansion}).

\subsubsection{Late expansion: No evidence for reacceleration}

\cite{bie11} reported an expansion curve with a hint of reacceleration at the latest epochs if the inner radius is also fit (in that case, the shell width increases at the late epochs; see their Fig. 4). However, the shell size, $R_o$, and the size ratio, $R_i/R_o$ (the complementary of the relative shell width), are anticorrelated in the model fitting (see, for instance, our Fig. \ref{Chi2DFig}). This means that fitting wider shell widths may imply fitting larger shell sizes, not because of a true increase in the shell size and/or width, but due to limitations in the estimate of the model parameters.

It is possible that Fig. 4 of \cite{bie11} is indeed reflecting such a behavior. Their last 2$-$4 epochs are the most affected by noise and, depending on the UV tapering used by the authors, estimating too wide a shell width would translate into a spurious signal of reacceleration in the expansion for these few epochs. Actually, looking at their Fig. 4, the expansion resulting from a fit with a fixed shell width (blue line) does not show such a signal of reacceleration.

In our analysis (Fig. \ref{ExpansionFig}, top panel), there is no hint of any reacceleration, unless we use narrow UV tapers (Fig. \ref{ExpansionUVTaperFig}, top panels), which heavily affect the fit, as is discussed in Sect. \ref{MethodSec}. Even though our fit $R_o$ follow the $\propto t^m$ law even at the latest times, it is not the case for $R_i$ (Fig. \ref{fig:RiExpansion}, left panel), which falls below a uniform expansion model at the latest epochs, which hence indicates a small widening of the shell.

\subsubsection{Wavelength effects in the shell}

\cite{Marcaide2009} reported an expansion curve of \SN93J with slightly different expansion indices at different frequencies. In particular, the authors found that, at late epochs, the observations at C band (5\,GHz) followed an expansion index of $m=0.788\pm0.015$, whereas at the L band (1.4\,GHz) the index was rather $m = 0.845\pm0.005$ (the same as that of C band in the earlier epochs). 

\cite{MartiVidal2011b} interpreted this wavelength effect in the expansion rate as due to the different ejecta opacity (which would decrease with time at a rate faster at the C band, compared to the L band). A lower ejecta opacity biases the shell size estimates toward lower values, and hence produces the lower $m$ value at C band. In other words, the wavelength effects in the expansion rate were an effect of a bias in the model fitting, rather related to ejecta opacity effects not considered in the shell model. 

If the interpretation of \cite{MartiVidal2011b} were correct, we should expect our Markov chains to produce estimates of $R_o$ that are actually independent of frequency, but to produce estimates of the inner radius, $R_i$, that are slightly higher at lower frequencies (in order to absorb the effects of the larger opacity from the ejecta). We notice, though, that physical scenarios different than that of the jet opacity can also reproduce the observational signature of a lower $R_i$ at higher frequencies, as it is discussed in Sect \ref{DiscussSec2}. 

\begin{figure}[ht!]
\includegraphics[width=9cm]{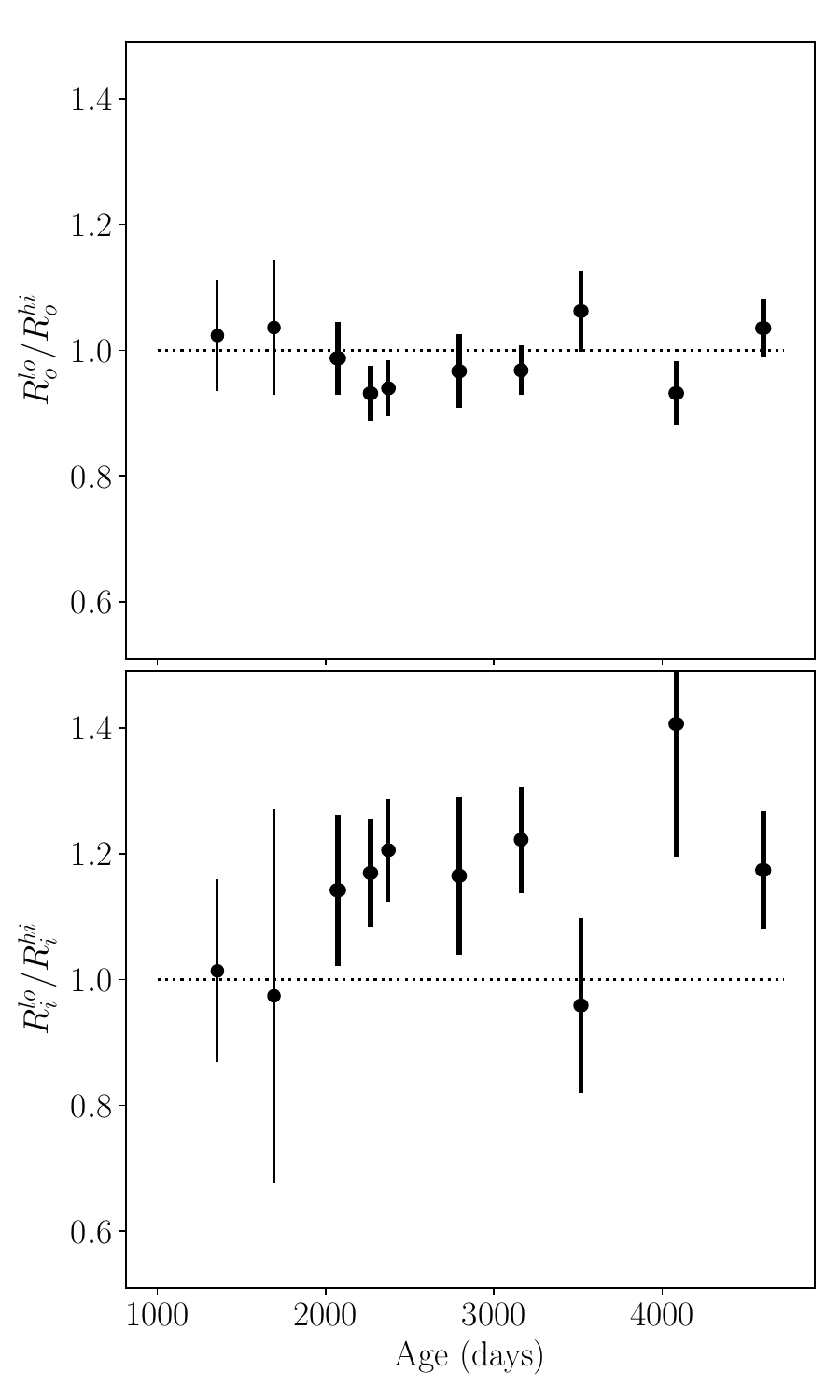}
\caption{%\LEt{***The first sentence of each figure caption or legend should be a descriptive title of the figure as a whole, written in a telegraphic style with the initial article (the, a, an) omitted. The text that follows should be written in complete sentences. If individual panels are described, this is repeated (another telegraphic description followed by full sentences).}
Frequency effects in the expansion curve.
{\bf Top} Ratio of outer shell sizes between data at low frequencies (L and S bands) and data at high frequencies (C and X bands) as a function of time. {\bf Bottom} Ratio of inner shell sizes between low and high frequencies. All pairs of epochs at different frequencies within 20 days of observation have been combined in this figure.}
\label{SizeWithNuFig}
\end{figure}

We selected a subset of data consisting of the epochs that were observed at C and L bands within a difference of 20\,days in the supernova age (to be able to assume a negligible change in the shell between observations). In Fig. \ref{SizeWithNuFig}, we show the ratios of outer shell sizes (top panel) and inner shell sizes (bottom panel) between frequencies (L band over C band), as a function of time. Indeed, as it can be seen in the figure, the ratios of outer sizes are all compatible with unity within the uncertainties, whereas for the ratios of inner sizes there is strong evidence of systematically larger $R_i$ at L band beyond day 1500 (the only exception is the epoch at day 3511). These results confirm the interpretation of the wavelength effects reported in \cite{Marcaide2009} as not related to $R_o$, but rather to effects from the inner side of the shell. %According to \cite{MartiVidal2011b}, the frequency-dependent ejecta opacity would produce such effects.

The weighted average of the $R_i$ ratios between the L and C bands (shown in Fig. \ref{SizeWithNuFig}) is $R_i^{lo}/R_i^{hi} = (1.17 \pm 0.02)$, which is more than 8$\sigma$ higher than unity. The Markov chains thus show strong evidence of wavelength effects in the supernova brightness structure, as was reported previously in \cite{Marcaide2009} and \cite{MartiVidal2011a, MartiVidal2011b}

\subsubsection{Lessons learned for VLBI observations of supernovae}

The analysis we have carried out in this paper on the VLBI data of SN 1993J, which spans almost 20 years, leads to a number of important results, whose implications should be taken into account when interpreting VLBI observations (and drawing conclusions) for this and other sources. These conclusions have implications that go far beyond the field of SNe, and are of general application, but we focus here on the case of SNe and, especifically, SN\,1993J.

First, we find that there are inherent, unavoidable issues in the modeling of VLBI data. 
Namely, VLBI array calibration is always imperfect, which can often lead to systematic effects in the analysis. Similarly, the choice of the deconvolution and/or modeling parameters may also systematically bias the final results. Those aspects are particularly relevant for SNe, which are relatively radio-faint, so that both calibration and image reconstruction algorithms can introduce artefacts in the final images, especially in cases of low S/N. 
Second, we also find that there is a strong coupling or degeneracy among some of the parameters. %\LEt{***Slashes are reserved to denote ratios, instrument pairings, and wavelength ranges (e.g., optical/UV), and for use in equations. The use of "and/or" is acceptable. Kindly rephrase here and where needed.} 
Without disentangling these degeneracies, or without a proper understanding of them, different results may be obtained with the same data, strongly depending on the calibration and image reconstruction algorithms. In turn, this may lead to different physical interpretations of the origin of the radio emission in SNe. In the particular case of SN\,1993J, a clear example is the coupling between shell size, relative shell width, radial-intensity profile, and absorption effects from the inner ejecta.

Our main lesson out of this paper is that, in order to fully understand the coupling of the model parameters in VLBI observations, and to properly tackle instrumental issues (array calibration) and (nonoptimal) choices of decovolution and imaging parameters, one needs to apply a complete sampling of the parameter space that defines the source structure and the basic antenna calibration quantities (antenna gains). In this sense, the use of techniques based on Monte Carlo Markov chains seems to be a good approach.

\subsection{Implications for models of shock expansion}
\label{DiscussSec2}

The power law decrease in the radio light curves is likely due to the self-similar structure of the inter-shock region in SN\,1993J. However, the achromatic break at around 3000\,days is not well described as a transition to another power law phase \citep{wei07}, which indicates the beginning of a more complex structure of the emission region. As was mentioned in the introduction, this transition may be caused by the reverse shock reaching the interface of the hydrogen and helium shells. A detailed discussion of this phase requires the numerical solution of the time-dependent hydrodynamical equations, which is beyond the scope of the present paper. Instead, it is interesting to undertake a comparison with the radio and X-ray structure of Cassiopeia\,A, which was also a type SN\,IIb. This supernova remnant is in a much later evolutionary stage, where the momentum input from the ejecta is unlikely to be high enough to maintain a self-similar structure of the inter-shock region. Hence, deviations from a self-similar structure may have characteristics in common in these two supernovae.

\cite{got01} show that in its inner, dominant emission region, the radial distribution of the radio intensity in Cassiopeia\,A follows the thermal X-ray intensity quite closely, while in the weaker, outer region the radio intensity correlates well with a hard, almost emission-line-free X-ray component. The extent of the emission regions is larger than expected in a self-similar situation. \cite{got01} noted that the swept-up circumstellar mass is roughly equal to the ejecta mass. This indicates that the supernova remnant is actually close to entering the Sedov phase and they proposed that the lagging behind of the reverse shock is due to the resulting rapid decline of the momentum input from the ejecta. One may note that there is also a third component, situated interior to the major emission region, which is dominated by thermal X-ray emission and which has indications of weak radio emission with a similar radial distribution.

\cite{got01} associated the weak outer region with the forward shock and placed the reverse shock at the steeply rising part of the dominant emission region. Although the identification of the forward shock is likely correct, there are a few implications of their position for the reverse shock that are harder to justify. If the supernova remnant is in a transition to the Sedov phase, the pressure and energy density %\LEt{***Slashes are reserved to denote ratios, instrument pairings, and wavelength ranges (e.g., optical/UV), and for use in equations. The use of "and/or" is acceptable. Kindly rephrase here and where needed.} 
behind the reverse shock should be substantially lower than behind the forward shock, and hence it is not clear how such an environment could produce most of the emission in the inter-shock region. Furthermore, this would place the third emission region just upstream of the reverse shock. However, this region has strong helium-like Si lines, which suggests that, instead, it corresponds to the downstream region of the reverse shock. This would then imply that the main emission is associated with the region just above the contact discontinuity.

Although in SN\,1993J the swept-up circumstellar mass is only a small fraction of the total ejecta mass, the transition of the reverse shock from the hydrogen to the helium shell is expected to affect the dynamics in a way similar to that in Cassiopeia\,A, since a flatter density distribution associated with such a transition would substantially decrease the momentum input. However, the effects are likely to be less dramatic than for Cassiopeia\,A, where the momentum input is about to cease; for example, the relative strength of the emission from the third component is expected to be somewhat stronger in SN\,1993J, since the inflow of momentum through the reverse shock just transitions into another ejecta structure rather than stopping altogether.

The modeling of the radio intensity in SN\,1993J made in this paper makes a uniform distribution of radio emission unlikely. Although the overall structure is consistent with a spherical source, the radial intensity distribution is not uniform. Instead, the intensity appears associated with the contact discontinuity, since the peak (of the Gaussian radial profile) is close to or just above the region where the latter is expected to be located. Furthermore, the model gave a value for the inner radius corresponding, roughly, to half the distance to the outer radius, which strongly indicates a non-self-similar structure of the emission region. Such a position of the inner radius is also consistent with the relocation of the reverse shock in Cassiopeia\,A, as was discussed above. The deduced properties of the radio emission suggest that at least the amplification of the magnetic field is due to processes related to the contact discontinuity; for example, the Rayleigh-Taylor instability \citep{bjo15}.

It is interesting to note that in SN\,1993J, the intensity peaks at a distance corresponding, roughly, to 0.8 of the outer radius. This is close to the value observed in Cassiopeia\,A ($\approx 0.7$). The velocity of the reverse shock is determined by the instantaneous influx of momentum from the ejecta, while the dynamics of the forward shock depends mainly on the accumulated momentum during the lifetime of the source. Hence, when a sudden drop of momentum input occurs at the reverse shock, its velocity is expected to respond quickly. The velocity of the forward shock is less affected and should decrease over a much longer timescale. This is particularly the case for SN\,1993J, since its velocity before the transition is already quite close to that for the Sedov phase. The responds of the contact discontinuity should be somewhere in between that of the reverse and forward shocks. Since Cassiopeia\,A has been in a non-self-similar state longer than SN\,1993J, one would then expect its contact discontinuity to be at a relative radius smaller than for SN\,1993J.

It is seen in Fig. \ref{SizeWithNuFig} that, starting at around 2000\,days, there is a tendency for the inner radius to be larger at low frequencies than at high frequencies. \cite{MartiVidal2011b} have suggested that this is an opacity effect; the decreasing free-free optical depth in the region interior to the reverse shock allows an increasing fraction of the emission from the back side of the source to add to the emission from the front side. This additional contribution to the emission would cause the light curves to flatten. This expectation is opposite to observations, which instead show rapidly steepening light curves. Hence, such an explanation is unlikely. However, a weakening of the reverse shock could cause a similar effect; the free-free absorption in the shocked ejecta would decrease as the density decreases and the temperature increases as a result of the lagging behind of the reverse shock. Such a scenario requires that a significant fraction of the emission comes from the shocked ejecta and that this emission is at least partly absorbed before the reverse shock starts to weaken. Although the first assumption is consistent with observations, the second is unlikely to be met, since the standard model for SN\,1993J implies a free-free optical depth much smaller than unity. 

Another possibility is  the Razin effect. This is not an opacity effect but affects the emission process directly. The thermal plasma causes the phase velocity of electromagnetic waves to become larger than the speed of light in vacuum. This leads to a dramatic decrease in the relativistic boosting of the synchrotron radiation below a frequency $\nu_{\rm R} = 20 n_{\rm e}/B$, where $n_{\rm e}$ is the density of thermal electrons \citep{g/s65}. Although various assumptions regarding the mass-loss rate of the progenitor star have been made \citep[see][for a discussion]{bjo15}, $n_{\rm e} \sim 10^5$ is likely for the density behind the reverse shock just before the break in the light curves. An estimate of the value of $B$ at this time can be obtained by fitting a homogenous source model to the observations. This gives $B\approx 3\times 10^{-2}$, which, in turn, implies $\nu_{\rm R} \sim 10^8$. 

Hence, in order for the Razin effect to account for the frequency dependence of the location of the inner radius, two things need to be fulfilled; namely, the weakening of the reverse shock should result in an average increase in the value of $\nu_{\rm R}$ by an order of magnitude and, furthermore, its value should increase continuously when going from the contact discontinuity toward the reverse shock.

If the emission structure of SN\,1993J is similar to that of Cassiopeia\,A, the magnetic field is likely amplified by turbulence. This turbulence is thought to be caused by the Rayleigh-Taylor instability emanating from the contact discontinuity, which, in turn, is driven by the strength of the reverse shock. As the reverse shock weakens, one may expect the turbulence, and hence the strength of the magnetic field, to decline from the contact discontinuity toward the reverse shock. 

However, as the reverse shock weakens, the electron density as well as the strength of the turbulence are expected to decrease. The density is directly related to the observed X-ray emission, while the synchrotron emission depends on the magnetic field strength. Hence, a larger value for $\nu_{\rm R}$ implies that the observed radio emission should decline more rapidly than the X-ray emission after the break. This is in accord with observations, which show that the X-ray light curves fall off more gradually than the radio ones do.

\section{Summary and conclusions}
\label{ConclusionSec}

SN\,1993J has been observed with VLBI for almost 20 years. From the first detection of its spherical shell-like structure \citep{MarcaideNat} and expansion \citep{MarcaideSci}, many publications have reported detailed results about its changing shock morphology and deceleration \citep[e.g.,][]{Marcaide1997,Bietenholz2001,Bartel2002,Marcaide2009,MartiVidal2011a}. Even though many of the conclusions from the different teams agree with each other (e.g., deceleration in the expansion curve), there are some important points where the different analyses diverge (e.g., relative shell width, wavelength effects in the shell structure and late evolution). This is of special interest, since these differing results lead to inconsistent physical interpretations about the details of the hydrodynamics and the shock radio emission.

We have carried out a new analysis on the complete set of VLBI observations of SN\,1993J (including one last epoch at 1.4\,GHz that was not reported previously in a refereed publication) that accounts for different instrumental and source-intrinsic effects, with the goal of obtaining information about SN\,1993J that is as robust and model-independent as possible. Our method uses the posterior probability distribution sampling based on Markov chains, which allows us to quantify the error budget of the different shell parameters and their complex correlations. 
Given the different results reported for this supernova by different teams,  
it appears clear that calibration and modeling of VLBI data, especially of sources with low brightness, 
can be easily biased. In particular, this is true if the models used are too restrictive and do not take into account possible effects related to instrumental calibration or departures between the model assumptions and the true brightness distribution of the source. In this sense, the use of a complete (instrument + source) consistent modeling, based on a full sampling of the posterior parameter space, seems to be a robust approach to characterize this kind of observation. 

Our conclusions regarding the radio emission structure in SN\,1993J are as follows:

\begin{enumerate}
    \item The main result is that the radial intensity distribution is not uniform. It peaks around or just above where the contact discontinuity is expected to be located. 
    \item The achromatic break in the radio light curves at around 3000 days is accompanied by an increasing deceleration of the reverse shock. In contrast, the behavior of the forward shock is seemingly unaffected by this dynamical transition.
    \item There are two indications that at least the magnetic field amplification is due to the Rayleigh-Taylor instability; namely, the location of the peak in the radial intensity distribution and the association of the achromatic break in the radio light curves with a weakening of the reverse shock.
    \item It is argued that the dynamical change at around 3000 days is due to the reverse shock reaching the transition region between the hydrogen and helium shells of the ejecta.
    \item The deduced properties of the structure of the radio emission region in SN\,1993J are quite similar to the spatially well-resolved supernova remnant Cassiopeia A. 
\end{enumerate}

\begin{acknowledgement}
This work has been partially supported by the Generalitat Valenciana GenT Project CIDEGENT/2018/021 and by the MICINN Research Projects PID2019-108995GB-C22 and PID2022-140888NB-C22. IMV acknowledges support from the Astrophysics and High Energy Physics programme by MCIN, with funding from European Union NextGenerationEU (PRTR-C17I1) and the Generalitat Valenciana through grant ASFAE/2022/018.
MPT acknowledges financial support 
from the Severo Ochoa grant CEX2021-001131-S and from the National grant 
PID2020-117404GB-C21, funded by MCIU/AEI/ 10.13039/501100011033.
\end{acknowledgement}

\newpage

\begin{appendix}

\section{Table of epochs and MCMC statistics}
\label{app:Table}

Table \ref{tab:Epochs} gives a list of the VLBI epochs included in this work, together with average values (and their standard deviations) of the posterior parameter distributions from the homogeneous shell model, using a visibility weighting with $T=30$ (see Eq. \ref{taperEq}). Notice that the distributions are non-Gaussian, so the quantities given in Table \ref{tab:Epochs} should be interpreted carefully.

\begin{table}
\caption{VLBI epochs of SN\,1993J and posterior statistics for a uniform shell model.}
\label{tab:Epochs}
\centering
\begin{tabular}{c r c c c}
\hline \hline
Project & Age~~ & Frequency & $R_o$ & $R_i$ \\
  Code  & (days) & (GHz)   & (mas) & (mas) \\
\hline
GR010 & 687 & 8.43 & 1.57 $\pm$ 0.05 & 1.04 $\pm$ 0.06 \\
GM21C & 697 & 5.00 & 1.51 $\pm$ 0.19 & 1.12 $\pm$ 0.21 \\
GR010 & 774 & 5.00 & 1.66 $\pm$ 0.10 & 1.23 $\pm$ 0.13 \\
GM024 & 774 & 5.00 & 1.69 $\pm$ 0.18 & 1.24 $\pm$ 0.25 \\
GR010 & 774 & 8.42 & 1.75 $\pm$ 0.31 & 1.21 $\pm$ 0.31 \\
GR010 & 873 & 5.00 & 1.91 $\pm$ 0.10 & 1.23 $\pm$ 0.14 \\
GR010 & 873 & 8.42 & 1.90 $\pm$ 0.14 & 1.19 $\pm$ 0.21 \\
GM025 & 917 & 5.00 & 1.95 $\pm$ 0.21 & 1.19 $\pm$ 0.40 \\
GR012 & 996 & 8.39 & 2.07 $\pm$ 0.12 & 1.20 $\pm$ 0.25 \\
BM057 & 1096 & 5.00 & 2.21 $\pm$ 0.29 & 1.24 $\pm$ 0.35 \\
GR012 & 1107 & 5.00 & 2.27 $\pm$ 0.14 & 1.30 $\pm$ 0.19 \\
GR012 & 1107 & 8.43 & 2.57 $\pm$ 0.08 & 1.09 $\pm$ 0.16 \\
GM027 & 1178 & 4.96 & 2.46 $\pm$ 0.18 & 1.66 $\pm$ 0.24 \\
BR041 & 1253 & 4.99 & 2.31 $\pm$ 0.09 & 2.04 $\pm$ 0.10 \\
BR041 & 1253 & 8.42 & 2.47 $\pm$ 0.12 & 1.70 $\pm$ 0.15 \\
GM027 & 1305 & 4.96 & 2.37 $\pm$ 0.08 & 2.13 $\pm$ 0.09 \\
BR044 & 1357 & 2.27 & 2.61 $\pm$ 0.17 & 2.05 $\pm$ 0.23 \\
BR044 & 1357 & 4.97 & 2.58 $\pm$ 0.15 & 2.03 $\pm$ 0.19 \\
BR044 & 1357 & 8.41 & 2.52 $\pm$ 0.12 & 2.02 $\pm$ 0.14 \\
GM027 & 1431 & 4.96 & 2.61 $\pm$ 0.11 & 2.28 $\pm$ 0.13 \\
GR015 & 1532 & 4.99 & 2.90 $\pm$ 0.15 & 2.10 $\pm$ 0.18 \\
GR015 & 1532 & 8.42 & 3.00 $\pm$ 0.12 & 2.05 $\pm$ 0.14 \\
GM030 & 1639 & 4.96 & 2.81 $\pm$ 0.09 & 2.54 $\pm$ 0.10 \\
GR017 & 1693 & 2.26 & 3.33 $\pm$ 0.33 & 2.08 $\pm$ 0.57 \\
GR017 & 1693 & 4.98 & 3.04 $\pm$ 0.15 & 2.34 $\pm$ 0.19 \\
GR017 & 1693 & 8.41 & 3.27 $\pm$ 0.06 & 2.06 $\pm$ 0.08 \\
GM030 & 1789 & 4.96 & 3.18 $\pm$ 0.12 & 2.49 $\pm$ 0.15 \\
GM035 & 1890 & 4.96 & 3.35 $\pm$ 0.05 & 2.58 $\pm$ 0.06 \\
GB027 & 1892 & 4.98 & 3.27 $\pm$ 0.13 & 2.65 $\pm$ 0.14 \\
GB027 & 1893 & 8.41 & 3.55 $\pm$ 0.06 & 2.21 $\pm$ 0.06 \\
GB033 & 2064 & 4.98 & 3.67 $\pm$ 0.22 & 2.85 $\pm$ 0.27 \\
GM035 & 2067 & 4.96 & 3.77 $\pm$ 0.04 & 2.60 $\pm$ 0.05 \\
GM035 & 2074 & 1.63 & 3.70 $\pm$ 0.16 & 3.02 $\pm$ 0.22 \\
GB033 & 2081 & 2.26 & 4.17 $\pm$ 0.34 & 2.26 $\pm$ 0.60 \\
GB033 & 2081 & 8.41 & 3.64 $\pm$ 0.11 & 2.69 $\pm$ 0.12 \\
GB033 & 2261 & 1.66 & 3.76 $\pm$ 0.13 & 3.40 $\pm$ 0.17 \\
GM035 & 2266 & 4.96 & 4.01 $\pm$ 0.06 & 2.94 $\pm$ 0.07 \\
GB033 & 2272 & 4.98 & 3.90 $\pm$ 0.28 & 2.92 $\pm$ 0.39 \\
GM035 & 2370 & 4.96 & 4.20 $\pm$ 0.05 & 2.92 $\pm$ 0.07 \\
GB035 & 2376 & 1.66 & 3.95 $\pm$ 0.19 & 3.52 $\pm$ 0.22 \\
GB035 & 2432 & 4.98 & 4.27 $\pm$ 0.23 & 2.84 $\pm$ 0.34 \\
GB034 & 2525 & 8.41 & 4.91 $\pm$ 0.38 & 2.38 $\pm$ 0.66 \\
GM035 & 2628 & 4.96 & 4.59 $\pm$ 0.09 & 3.05 $\pm$ 0.09 \\
GB038 & 2787 & 2.26 & 4.97 $\pm$ 0.38 & 3.22 $\pm$ 0.59 \\
GB038 & 2787 & 8.41 & 5.13 $\pm$ 0.19 & 2.92 $\pm$ 0.22 \\
GM040 & 2795 & 1.63 & 4.58 $\pm$ 0.19 & 4.04 $\pm$ 0.23 \\
GM040 & 2799 & 4.96 & 4.93 $\pm$ 0.08 & 3.25 $\pm$ 0.09 \\
GM040 & 2881 & 4.96 & 5.00 $\pm$ 0.04 & 3.51 $\pm$ 0.05 \\
GB042 & 2996 & 4.99 & 5.07 $\pm$ 0.10 & 3.44 $\pm$ 0.09 \\
GM046 & 3158 & 4.96 & 5.54 $\pm$ 0.08 & 3.26 $\pm$ 0.09 \\
GB042 & 3165 & 1.66 & 5.37 $\pm$ 0.18 & 3.98 $\pm$ 0.22 \\
GB042 & 3345 & 4.99 & 5.51 $\pm$ 0.10 & 3.65 $\pm$ 0.11 \\
GM048 & 3512 & 4.96 & 6.03 $\pm$ 0.25 & 3.89 $\pm$ 0.33 \\
GM048 & 3522 & 1.63 & 6.53 $\pm$ 0.45 & 3.48 $\pm$ 0.72 \\
GB046 & 3718 & 4.98 & 5.50 $\pm$ 0.32 & 4.39 $\pm$ 0.39 \\
GM048 & 3868 & 4.96 & 6.47 $\pm$ 0.15 & 3.87 $\pm$ 0.23 \\
GB049 & 4077 & 4.98 & 6.95 $\pm$ 0.24 & 3.39 $\pm$ 0.44 \\
GB049 & 4089 & 1.66 & 6.50 $\pm$ 0.25 & 4.76 $\pm$ 0.30 \\
GM055 & 4236 & 4.96 & 6.45 $\pm$ 0.33 & 4.50 $\pm$ 0.42 \\
GB049 & 4459 & 4.98 & 6.67 $\pm$ 0.13 & 4.48 $\pm$ 0.10 \\
GM057 & 4592 & 4.99 & 6.54 $\pm$ 0.24 & 4.33 $\pm$ 0.28 \\
GM057 & 4607 & 1.66 & 6.82 $\pm$ 0.16 & 5.07 $\pm$ 0.20 \\
GB058 & 4829 & 4.98 & 7.23 $\pm$ 0.27 & 4.75 $\pm$ 0.30 \\
GB062 & 5334 & 4.98 & 8.32 $\pm$ 0.43 & 4.39 $\pm$ 1.07 \\
GB070 & 6187 & 1.66 & 9.16 $\pm$ 0.17 & 5.23 $\pm$ 0.22 \\
\hline
\end{tabular}

\end{table}

\end{appendix}

\end{document}